\documentclass[aps,pre,twocolumn,showpacs,superscriptaddress]{revtex4}
\usepackage{amsfonts,amssymb,amsmath,times}
\usepackage{graphicx}
\usepackage{bm}
\usepackage{enumerate}
\usepackage{color}
\usepackage{amsmath,amssymb}
\usepackage{mathrsfs}
\usepackage{rotating}
\usepackage{soul}
\usepackage{cancel}

\begin{document}

\title{Disentangling different types of El Ni{\~{n}}o episodes by evolving climate network analysis}

\author{A.~Radebach}
 \affiliation{Potsdam Institute for Climate Impact Research, P.\,O.~Box 60\,12\,03, 14412 Potsdam, Germany}
 \affiliation{Department of Physics, Humboldt University Berlin, Newtonstra{\ss}e 15, 12489 Berlin, Germany}
 \affiliation{Mercator Research Institute on Global Commons and Climate Change, Torgauer Stra{\ss}e 12-15, 10829 Berlin, Germany}
 \affiliation{Macroeconomic Sustainability Assessment, Economics of Climate Change, Technical University Berlin, Stra{\ss}e des 17.~Juni 145, 10623 Berlin, Germany}
\author{R.\,V.~Donner} \email[Corresponding author: ]{reik.donner@pik-potsdam.de}
 \affiliation{Potsdam Institute for Climate Impact Research, P.\,O.~Box 60\,12\,03, 14412 Potsdam, Germany}
 \affiliation{Max Planck Institute for Biogeochemistry, Hans-Kn\"oll-Stra{\ss}e 10, 07745 Jena, Germany}
\author{J.~Runge}
 \affiliation{Potsdam Institute for Climate Impact Research, P.\,O.~Box 60\,12\,03, 14412 Potsdam, Germany}
 \affiliation{Department of Physics, Humboldt University Berlin, Newtonstra{\ss}e 15, 12489 Berlin, Germany}
\author{J.\,F.~Donges}
 \affiliation{Potsdam Institute for Climate Impact Research, P.\,O.~Box 60\,12\,03, 14412 Potsdam, Germany}
 \affiliation{Department of Physics, Humboldt University Berlin, Newtonstra{\ss}e 15, 12489 Berlin, Germany}
 \affiliation{Stockholm Resilience Centre, Stockholm University, Kr\"aftriket 2B, 11419 Stockholm, Sweden}
\author{J.~Kurths}
 \affiliation{Potsdam Institute for Climate Impact Research, P.\,O.~Box 60\,12\,03, 14412 Potsdam, Germany}
 \affiliation{Department of Physics, Humboldt University Berlin, Newtonstra{\ss}e 15, 12489 Berlin, Germany}
 \affiliation{Institute for Complex Systems and Mathematical Biology, University of Aberdeen, Aberdeen AB243UE, United Kingdom}

\date{\today}

\begin{abstract}
Complex network theory provides a powerful toolbox for studying the structure of statistical interrelationships between multiple time series in various scientific disciplines. In this work, we apply the recently proposed climate network approach for characterizing the evolving correlation structure of the Earth's climate system based on reanalysis data of surface air temperatures. We provide a detailed study on the temporal variability of several global climate network characteristics. Based on a simple conceptual view on red climate networks (i.e., networks with a comparably low number of edges), we give a thorough interpretation of our evolving climate network characteristics, which allows a functional discrimination between recently recognized different types of El Ni{\~{n}}o episodes. Our analysis provides deep insights into the Earth's climate system, particularly its global response to strong volcanic eruptions and large-scale impacts of different phases of the El Ni{\~{n}}o Southern Oscillation (ENSO). 
\end{abstract}

\pacs{92.60.Ry, 92.10.am, 89.75.Hc, 05.45.Tp}

\maketitle

\section{Introduction} \label{sec:introduction}

During the last years, complex network theory~\cite{Albert2002,Newman2003,Boccaletti2006} has found wide use not only in the social sciences, engineering, and biology, but also in Earth and environmental sciences. Pioneering work on fundamental aspects of many real-world complex networks has triggered an enormous interest in applying graph-theoretical concepts for the characterization of complex geophysical systems. Among others, prominent examples include applications in hydrology~\cite{Zaliapin2010}, seismology~\cite{Abe2004,Baiesi2004,Davidsen2008,Jimenez2009}, soil sciences~\cite{Vogel2001,Santiago2008,Mooney2009}, and geoscientific time series analysis~\cite{Elsner2009,Donges2011NPG,Donges2011PNAS,Donner2012AG,Telesca2012,Pierini2012,Feldhoff2012}. Recently, also climatologists started to discover the instruments of complex network theory \cite{Tsonis2004,Tsonis2006}. Having lead to novel insights into the climate system, this promising new branch of climate science is on its way to refine and consolidate its tools \cite{Tsonis2008b,Tsonis2008c,Tsonis2010,Tsonis2012,Gozolchiani2008,Gozolchiani2011,Yamasaki2008,Yamasaki2009,Donner2008,Donges2009a,Donges2009b,Donges2011,Steinhaeuser2010a,Steinhaeuser2010b,Steinhaeuser2011,Steinhaeuser2010CIDU,Pelan2011,Kawale2011CIDU,Kawale2011SDM,Barreiro2011,Malik2010,Malik2011,Boers2013,Palus2011,Berezin2011,Guez2012,Ebert-Uphoff2012a,Ebert-Uphoff2012b,Feng2012,Carpi2012EPJB,Carpi2012Book,Rheinwalt2012,Fountalis2013,Steinhaeuser2013,Hlinka2013a,Hlinka2013b,Martin2013,vanderMheen2013,Tirabassi2013,Deza2013,Ludescher2013}.

In order to understand the functioning of the climate system, relevant underlying physical processes and their interactions have to be identified. For this purpose, a widely applicable approach is performing a careful statistical analysis of existing climate data and successively refining existing mathematical models. Here, we focus on the statistical aspect only. Traditionally, this problem has been addressed by methods from multivariate statistics, such as empirical orthogonal function (EOF) analysis and related techniques. In order to study spatio-temporal climate variability from a different perspective, the \textit{climate network} approach has been introduced for obtaining a spatially discretized representation of the spatially extended dynamical system ``climate'' based on significant statistical associations extracted from the multitude of entangled interactions in the original system \cite{Tsonis2004,Tsonis2012}. Thus, climate network analysis opens a new perspective on the Earth's complex climate system. 

The bridge from complex network theory to the climate system is based on two fundamental identifications. First, a distinct set of climatological time series obtained at fixed locations on the Earth are interpreted as vertices of the climate network. Second, relevant statistical associations between the time series are represented by the network's edges. The climate network resulting from this approach is then subject to certain well established {(}but still actively progressing{)} statistical methods originated in complex network theory \cite{Albert2002,Newman2003,Boccaletti2006,Costa2007}. While, as sketched above, this approach is a relatively young one in the climate context, the same structural identification is nowadays widely used in neuroscience, leading to so-called \textit{functional brain networks} \cite{McIntosh1994,Zhou2006,Schindler2008,Bullmore2009,Zamora2011} based on statistical associations between electromagnetic recordings at different parts of the brain.

Recent research on climate networks has either investigated several measures of the static network relying on the complete time span of observations \cite{Donges2009a,Malik2011,Palus2011} or considered the temporal variability of only one specific measure \cite{Yamasaki2008,Gozolchiani2008,Yamasaki2009,Gozolchiani2011,Guez2012,Berezin2011}. In this work, we combine these two approaches to analyze the time-evolution of the global climate system from a complex network perspective using a set of complementary network characteristics. A similar approach has been recently applied in the analysis of long-term variability in epileptic brain networks \cite{Kuhnert2010}. We emphasize that the approach of \textit{evolving networks} (i.e., complex network structures representing the system's state within several consecutive windows in time) as used in this work is conceptually related with, but distinctively different from \textit{temporal networks} \cite{Holme2012}. Notably, the concept of temporal networks explicitly mixes topological and temporal information, whereas both are clearly separated in the present study.

In this paper, we present several methodological improvements with respect to previous works on climate networks, such as an alternative type of spatial grid for the network construction, which avoids distortions of the climate network's properties due to the grid geometry \cite{Donges2011}. Subsequently, we apply our modified approach to reanalysis data of surface air temperature around the globe, spanning the time period between 1948 and 2009. The meaning of characteristic graph properties of the climate network in terms of the underlying physical system as well as their temporal variability when obtained from running windows in time are systematically studied and discussed in the context of known large-scale climate events such as El Ni{\~{n}}o episodes or strong volcanic eruptions.

This paper is organized as follows: In Sect.~\ref{sec:methods}, we describe the data set used in this work. Afterwards, the construction and statistical description of climate networks is discussed in some detail. The results of an evolving climate network analysis are presented in Sect.~\ref{sec:results}. Subsequently, we demonstrate the robustness of our findings regarding various methodological options in Sect.~\ref{sec:robustness} and put them into a climatological context in Sect.~\ref{sec:discussion}. Finally, the main conclusions obtained from the presented work are summarized (Sect.~\ref{sec:conclusions}).

\section{Materials and methods}\label{sec:methods}

\subsection{Description of the data}

As stated above, climate networks are complex networks based on statistical associations between climatological time series obtained at several locations on the Earth. In this study, we use air temperatures obtained from the NCEP/NCAR reanalysis I data set~\cite{Kalnay1996}{,} which cover the time period 1948-2009 (i.e., $62$ years) with a daily resolution on an angularly regular {$2.5^\circ\times2.5^\circ$} grid. Specifically, we investigate air temperatures obtained at sigma level $0.995$ (i.e., the atmospheric height where $99.5\%$ of the surface air pressure is attained), shortly {referred} to as surface air temperatures (SAT) in the following.

The annual cycle of solar insolation is known to induce the leading-order variation of air temperatures. Since we are interested in dynamical interactions within the Earth's climate, this dominant externally triggered effect is not of interest. In order to properly remove the effect of seasonality from observational time series, a number of different methods may be used~\cite{Donner2008}. Here, we restrict ourselves to removing the long-term mean annual cycle (base period are the $62$ years of the record) from the observational data separately for all considered locations. For this purpose, we subtract the long-term mean values for each day of the year, a procedure known as phase averaging \cite{Donner2008}. Of course, the resulting first-order surface air temperature anomalies (SATA) only account for seasonality in the mean, while annual variations in higher-order statistical properties such as the variance are not removed. Moreover, interannual shifting of seasons {\cite{Thomson1995, Palus2005, Rybski2011}} is not considered. For technical reasons, all leap days are removed from the resulting time series, which has only negligible effects on the results as long as only lag-zero statistical associations between different sites are studied. Regarding non-zero lags, the corresponding effects are found to be statistically negligible as well.

\subsection{Climate network construction} \label{sec:construction}

\subsubsection{Identification of vertices (nodes)} 

\begin{figure}
\begin{center}
\includegraphics[width=0.485\textwidth]{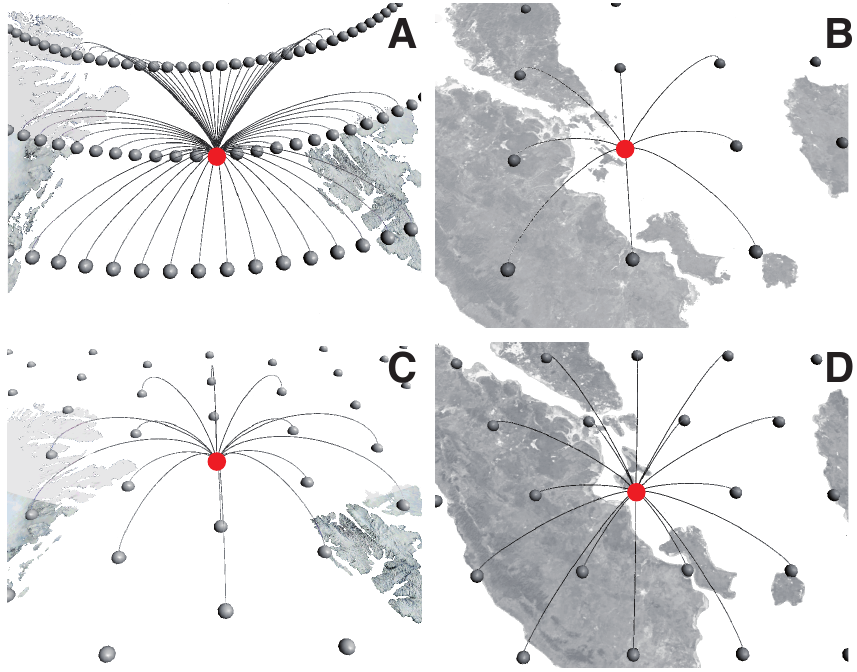}
\end{center}
\caption{(Color online) Geographical neighborhoods of (A,C) a high-latitude (North Atlantic between Svalbard and Northern Greenland) and (B,D) a low-latitude (close to Singapore) grid point given on (A,B) a standard (angularly regular) {$2.5^\circ\times2.5^\circ$} grid with $10{,}226$ grid points and (C,D) an icosahedral grid with $10{,}242$ grid points (i.e., $n=5$ completed refinement steps), respectively. The chosen grid points (red) are connected to all other grid points that are closer than $500$ kilometers. For the standard grid, the high-latitude grid point has $56$ geographical neighbors, whereas the one at low latitude has only $8$. In contrast, for the icosahedral grid, both grid points have $16$ neighbors each. Obviously the heterogeneous spatial distribution of grid points in the standard grid determines the huge difference between both neighborhoods, whereas the homogeneous spatial distribution of grid points of the icosahedral grid enhances the comparability of vertex properties in different regions of the Earth.}
\label{fig:grid}
\end{figure}

The first step in the construction of a climate network is the appropriate identification of vertices. For example, the locations at which the considered time series are available can be directly used as the spatial locations of network vertices. When operating with station data~\cite{Donner2008}, this leads to an irregular spatial distribution of vertices with a large variety of nearest-neighbor distances. However, even for reanalysis data sets or climate models, the arrangement of vertices in the published data sets is commonly only regular with respect to the difference angles in both longitudinal and latitudinal coordinates. This results in a significant spread in the actual spatial distances between neighboring vertices in low and high latitudes. Specifically, average inter-vertex distances are smaller close to the poles than in low-latitude regions (see Fig.~\ref{fig:grid}A,B). Such a heterogeneous distribution of vertices is known to induce severe distortions in the topological properties of spatially embedded networks~\cite{Heitzig2010,Bialonski2010}. Even more, for reanalysis data, information for high latitudes is typically provided with rather large uncertainty, since there are hardly any direct measurements that can be assimilated into the underlying climate model. As a consequence, there are many vertices with less reliable data in the polar regions.

In order to correct for the geometrically induced effects, Heitzig \textit{et~al.}~\cite{Heitzig2010} recently introduced a specific class of vertex-weighted network measures explicitly taking information on the spatial distribution of vertices and, hence, their neighborhood size into account. This concept can be understood as a sophisticated generalization of area-weighted connectivity measures previously studied in the context of climate network analysis \cite{Tsonis2006,Tsonis2008b,Donges2009a,Barreiro2011,Deza2013}. 

As an alternative approach, in this work we project the available spatially distributed SATA time series onto a different type of grid with a higher degree of homogeneity and isotropy on the sphere (i.e., a grid where the typical nearest-neighbor grid point distances as well as the numbers of neighboring vertices are the same almost everywhere) by means of interpolation \cite{Donges2011}. Specifically, we use a quasi-isotropic icosahedral grid \cite{Heikes1995} (see Fig.~\ref{fig:grid}C,D), which is constructed as follows: First, the vertices of an icosahedron are projected onto the sphere, yielding $12$ initial grid points with constant spacing. As anchor points defining the icosahedron, we use North and South Pole as well as a third point at $26.56^\circ$N, $0^\circ$E (the choice of this third reference point on the zero meridian is convenient, but arbitrary); all other initial grid points follow from symmetry. The same procedure applies to the edges of the icosahedron (forming $20$ equilateral triangles), which are also projected onto the sphere. Second, every projected triangle is partitioned into four smaller triangles with approximately the same area on the sphere by bisecting the projected edges. At each bisection point, a new vertex is introduced. This procedure of grid refinement is repeated as often as desired. The number of vertices grows as $N = 5\cdot 2^{2n+1} + 2$ with $n$ being the number of completed refinement steps, i.e., $N=42$, $162$, $642$, {$2{,}562$} and {$10{,}242$} for $n=1,\dots,5$. Conversion of the available data is performed using a standard bilinear interpolation scheme using the four angularly regular grid points of the quadrilateral surrounding the respective icosahedral grid point \cite{Jones1998}.

We emphasize that for the SATA data used in this study, the described spatial interpolation does not cause any considerable errors, since the SATA variability at geographically close points is typically very similar. However, interpolation can generally induce spurious correlations~\cite{Rehfeld2011}, which are not necessarily spatially homogeneous. Since the framework used in this paper is based on correlations between time series from different locations (see below), we cannot completely rule out a possible effect on the resulting climate network properties. Given the wide-spread use of such interpolation approaches in climate sciences, we conjecture that a possible bias (given its existence) can be widely neglected. A detailed examination of this point is, however, beyond the scope of the present study.

\subsubsection{Identification of edges (links)} \label{sec:edges}

Having thus defined the vertices of the climate network, in a second step, the corresponding connectivity is established. This step requires two basic ingredients: the selection of a pair{-}wise measure of statistical association between time series obtained at each grid point (vertex), and the definition of an appropriate threshold criterion determining which of these associations are statistically relevant. Specific association measures previously used for climate network construction include the linear (Bravais-Pearson) correlation coefficicent \cite{Tsonis2004}, (cross-) mutual information \cite{Donges2009a,Donges2009b}, a phase synchronization index based on the normalized Shannon entropy of the associated phase difference time series \cite{Yamasaki2009}, the (cross-) mutual information of order patterns \cite{Runge2010,Barreiro2011,Tirabassi2013,Deza2013}, event synchronization \cite{Malik2010,Malik2011,Boers2013}, transfer entropy \cite{Hlinka2013b}, or graphical models for identifying ``causal'' climate networks \cite{Ebert-Uphoff2012a,Ebert-Uphoff2012b,Runge2012,Runge2012b}. We refer to the corresponding references for details. Of course, other association measures could be used here as well. 

In all cases, the resulting matrix of normalized pair{-}wise statistical associations, e.g., cross-correlation coefficients (here within a given time window) is considered as the weight matrix $\mathbf{W}=(W_{ij})$ of a fully connected weighted graph. In order to obtain a climate network representation (as a simple unweighted graph), thresholding is applied to this matrix to infer the climate network's adjacency (connectivity) matrix $\mathbf{A}=(A_{ij})$ defined as
\begin{equation}
A_{ij}=\Theta(W_{ij}-W_{ij}^*)-\delta_{ij}. 
\end{equation}
\noindent
Here, $W_{ij}^*$ is a threshold deciding whether or not the association between vertices $i$ and $j$ is considered statistically relevant, $\Theta(\cdot)$ is the Heaviside function, and $\delta_{ij}$ Kronecker's delta. In principle, this thresholding can be performed in two different ways:

\begin{enumerate}[(i)]

\item On the one hand, it is possible to \textit{locally} select an appropriate threshold separately for each pair of vertices \cite{Runge2010,Steinhaeuser2010b,Palus2011}, where the significance is determined independently by taking the individual time series' probability distribution and auto-covariance structure into account, for example, by means of AAFT surrogates \cite{Schreiber2000} or block-bootstrapping \cite{Davison1997}. In this spirit, local thresholding has the important conceptual advantage of representing only the statistically significant interrelationships with respect to some specific null model. 

\item On the other hand, the threshold can be defined \textit{globally}, i.e., $W_{ij}^*\equiv W^*$ \cite{Tsonis2004,Yamasaki2008,Donges2009b}. This can be achieved by considering a fixed quantile of the empirical distribution $p(W_{\bullet\bullet})$~\footnote{We adopt the notation $X_\bullet$ and $X_{\bullet\bullet}$ throughout the paper whenever we do not refer to a specific vertex, edge, or pair of vertices when discussing a certain property $X$.} of all weights $W_{ij}$ (e.g., determined by the significance of associations of a proper statistical model), which results in an edge density 
\begin{equation}
\rho=\left. \sum_{i=1}^{N-1} \sum_{j=i+1}^N A_{ij}\right/ \binom{N}{2}
\end{equation}
\noindent
(i.e., the fraction of possible edges realized in the network). Obviously, global thresholding is computationally by far less demanding than local thresholding and allows a more direct comparison of network patterns obtained at different parts of the globe. 

\end{enumerate}

Notably, both approaches are \emph{not} equivalent, since global thresholding can lead to spurious results in the presence of strong serial dependences (e.g., auto-correlations) in some individual time series \cite{Palus2011}. Nevertheless, in this work, we will restrict ourselves exclusively to global thresholding in order to reduce the computational efforts. Note that in general, thresholding results in a loss of information about the exact strengths of pair{-}wise associations. Hence, different thresholds represent different levels of considered association strength (or different significance levels in case of local thresholding) and result in different edge densities of the derived networks. Consequently, looking at climate networks with different edge densities highlights distinct aspects and intrinsic scales of the underlying association structure of the climate system.

\subsection{Network quantifiers} \label{sec:measures}

After having transformed the available climate data into a complex network representation, the next step is to characterize the resulting discrete structures. For this purpose, there is a large amount of statistical characteristics quantifying different aspects of network topology on both local (vertex or edge) and global scale~\cite{Albert2002,Newman2003,Boccaletti2006,Costa2007}. In recent research on climate networks, much attention has been spent on the probability distribution{s} and spatial patterns of vertex characteristics, such as degree  
\begin{equation}
k_i=\sum_{j=1}^N A_{ij},
\end{equation}
\noindent
area-weighted connectivity~\cite{Tsonis2006,Tsonis2008b,Barreiro2011,Palus2011}, or betweenness centrality~\cite{Donges2009a,Donges2009b,Palus2011}.

Besides such measures characterizing exclusively network \textit{topology} (connectivity), there are those quantifying certain aspects of the spatially embedded \textit{geometry} of the graph. These measures rely on the geographical distance matrix $\mathbf{d}=(d_{ij})$ which stores the shortest spatial distances (along geodesics on the sphere) between all pairs of vertices $i$ and $j$. Notable examples are the edge length distribution $p(d_{\bullet\bullet}|A_{\bullet\bullet}=1)$ -- in the following understood as referring to present edges -- and the maximal edge length per vertex
\begin{equation}
d_i^{\text{max}} = \max_j (d_{ij}|A_{ij}=1).
\end{equation}
The latter quantity allows identifying vertices possessing long-range connections (teleconnections).

In contrast to these local measures, in this work we are mostly interested in characterizing temporal changes of the climate network topology on a global scale, which primarily calls for the study of scalar-valued network characteristics evolving in time. Of course, one has to be aware of the fact that changes of such global characteristics always reflect changes at a local scale.

Temporal changes in climate networks have already been considered by different authors. Tsonis {and Swanson} \cite{Tsonis2008b} compared the number and geographic length distribution of edges as well as the spatial connectivity pattern for El Ni{\~{n}}o (EN) and La Ni{\~{n}}a (LN) years. They found that under EN conditions, the global climate network contains considerably fewer and geographically shorter edges when considering a fixed threshold $W^*$ for network construction. Using a more subtle approach, Yamasaki and co-workers \cite{Yamasaki2008,Gozolchiani2008,Yamasaki2009,Gozolchiani2011} confirmed a considerable global impact of El Ni{\~{n}}o on the climate network in terms of the appearance and disappearance of edges (``blinking links''). 

Here, we mainly focus on the time evolution of three global network characteristics{,} which are widely used in complex network research: 

\begin{enumerate}[(i)]

\item The average path length $\mathcal{L}$ \cite{Newman2003,Boccaletti2006} measures the mean shortest (geodesic) graph distance between all pairs of vertices in the network, i.e., the average smallest number of edges to be traversed to cover the distance between two randomly chosen vertices on the graph, 
\begin{equation}
\mathcal{L} = \frac{1}{N} \mathcal{L}_i \quad \mbox{with} \quad \mathcal{L}_i=\frac{1}{N-1}\sum_{j=1}^N \mathcal{L}_{ij},
\label{eq:apl}
\end{equation}
with $\mathcal{L}_{ij}$ denoting the length of the shortest path (i.e., the number of edges) between vertices $i$ and $j$, and $\mathcal{L}_{ii}=0$ by definition. Note that for ensembles of spatially embedded networks with the same edge density, transfer of connectivity between spatial scales (i.e., changes in the edge length distribution) can change the average path length. However, spatial redistribution of edges alone (i.e., even without transfer between spatial scales) can lead to similar changes in $\mathcal{L}$.

\item The network transitivity $\mathcal{T}$ \cite{Boccaletti2006} -- sometimes also referred to as the (Barrat-Weigt) clustering coefficient \cite{Barrat2000,Newman2003} -- characterizes the degree of transitivity in the connectivity relations in the network relative to the maximally possible value (or, put differently, the global density of closed ``triangles'' in the network): 
\begin{equation}
\mathcal{T}=\frac{\sum_{i,j,k=1}^N A_{ij} A_{ik} A_{jk}}{\sum_{i,j,k=1, j \neq k}^N A_{ij} A_{ik}}.
\label{eq:transitivity}
\end{equation}

\item The global (Watts-Strogatz) clustering coefficient $\mathcal{C}$ \cite{Watts1998} measures the average density of triangles centered at all vertices of a network, 
\begin{equation}
\mathcal{C}=\frac{1}{N} \sum_{i=1}^N \mathcal{C}_i \quad \mbox{with} \quad \mathcal{C}_i = \frac{\sum_{j,k=1}^N A_{ij} A_{ik} A_{jk}}{k_i(k_i-1)},
\label{eq:clustering}
\end{equation}
\noindent
where $\mathcal{C}_i$ is the local clustering coefficient of vertex $i$. $\mathcal{C}$ is conceptually related with, but distinct from $\mathcal{T}$ and actually captures a different property of the network under study. Particularly, $\mathcal{T}$ does not explicitly take the degree of each vertex into account, whereas $\mathcal{C}$ does.

\end{enumerate}

For spatially embedded networks such as climate networks, the possible ranges of the aforementioned global characteristics are often predetermined by the associated spatial constraints \cite{Bialonski2010,Bialonski2011,Dall2002,Donner2011EPJB,Donges2012PRE,Rheinwalt2012}, which calls for a careful interpretation of the corresponding results. For example, the small-world property (i.e., high global clustering coefficient and short average path length \cite{Watts1998}) common to many real-world networks can be induced by the spatial embedding alone \cite{Bialonski2010}.

\subsection{Characterization of graph dissimilarity} \label{sec:hamming}

In addition to the scalar network characteristics discussed above, for studying dynamical changes in climate network topology, it is useful to consider a measure for comparing two networks with the same set of vertices. This is traditionally achieved by the Hamming distance \cite{Hamming1950,Donges2009a}
\begin{equation}
\mathcal{H}(\mathscr{G},\mathscr{G}')=\left. \sum_{i=1}^{N-1} \sum_{j=i+1}^N \left| A_{ij}-A'_{ij}\right| \right/ \binom{N}{2},
\end{equation}
\noindent
where $\mathscr{G}$ and $\mathscr{G}'$ are the two graphs to be compared with adjacency matrices $\mathbf{A}$ and $\mathbf{A}'$, respectively. By definition, we have $\mathcal{H}=0$ for identical networks, and $\mathcal{H}=1$ for networks being inverse with respect to the presence and absence of edges. Note that $\mathcal{H}$ treats the combined presence and absence of edges in the two networks symmetrically, i.e., an inversion of both networks does not alter the result.

Trivially two networks with different numbers of edges always have $\mathcal{H}>0$. Hence, separating the corresponding effect from a ``real'' difference in the placement of (present) edges provides additional insights into network topology. Let us define
\begin{align*}
   a &= |\{ (i,j) | i<j \wedge A_{ij}=0 \wedge A'_{ij}=0\}| \\
   b &= |\{ (i,j) | i<j \wedge A_{ij}=1 \wedge A'_{ij}=0\}| \\
   c &= |\{ (i,j) | i<j \wedge A_{ij}=0 \wedge A'_{ij}=1\}| \\
   d &= |\{ (i,j) | i<j \wedge A_{ij}=1 \wedge A'_{ij}=1\}|,
\end{align*}
where $|S|$ is the number of elements of the set $S$, i.e., $a$ is the number of edges absent in both networks, $d$ is the number of edges present in both networks, and $b$ and $c$ refer to the respective numbers of edges present in exactly one of both networks. 
This implies $ \mathcal{H(\mathscr{G}, \mathscr{G}^\prime)}=\left. (b+c) \right/\binom{N}{2}$. Let $\rho = \left. (b+d)\right/\binom{N}{2}$ and $\rho' = \left. (c+d)\right/\binom{N}{2}$ be the edge densities of both networks. Without loss of generality, $\rho\geq \rho'$ (i.e., $b \geq c$). With the edge density difference $\Delta\rho := \left|\rho-\rho'\right| = \left. (b-c)\right/ \binom{N}{2}\geq 0$, we obtain
\begin{equation}
 \mathcal{H(\mathscr{G}, \mathscr{G}^\prime)} = \Delta\rho + 2 c\left/\binom{N}{2}\right.=\Delta\rho+ \mathcal{H^\star(\mathscr{G}, \mathscr{G}^\prime)},
\end{equation}
\noindent
i.\,e., $\mathcal{H^\star(\mathscr{G}, \mathscr{G}^\prime)}=2 c\left/\binom{N}{2}\right.$.
Recall that $c$ is the number of edges that are present in the network with the lower edge density but \textit{not} in the network with the higher edge density. For the mutual comparison of different climate networks the latter part, which we refer to as the \emph{corrected Hamming distance} 
\begin{equation}
\mathcal{H^\star(\mathscr{G}, \mathscr{G}^\prime)} = \mathcal{H(\mathscr{G}, \mathscr{G}^\prime)} - \Delta\rho , 
\end{equation}
\noindent
is of particular interest, since it measures the structural dissimilarity one would find for two networks of the same edge density, disregarding the effect of different numbers of edges. By definition, we have $0\leq \mathcal{H^\star(\mathscr{G}, \mathscr{G}^\prime)}\leq \mathcal{H(\mathscr{G}, \mathscr{G}^\prime)}\leq 1$, i.e., the same range as for the original Hamming distance. 

In the context of complex networks representing subsequent snapshots of the evolving network topology, i.e., $\mathscr{G}=\mathscr{G}_t$ and $\mathscr{G}'=\mathscr{G}_{t-\delta t}$ with $t$ denoting some time interval of interest and $\delta t$ being a fixed time increment, the Hamming distance 
\begin{equation}
\mathcal{H}_{t,t-\delta t}=\mathcal{H(\mathscr{G}, \mathscr{G}^\prime)} 
\label{def:hamming}
\end{equation}
\noindent
(as well as its counter-part corrected for the effect of different edge densities) can be interpreted as the relative change in connectivity between subsequent networks, i.e., a discrete ``network derivative'' given the direct analogy with the {classical} difference quotient. This viewpoint is of particular interest in the context of evolving climate networks, since strong differences between networks obtained for subsequent time intervals point to a (temporary) global-scale instability of the spatial interdependence structure of the considered climatological observable.

\section{Results} \label{sec:results}

\subsection{Methodological setting}\label{sec:timdep}

In order to study the signatures of annual- to decadal-scale variability in the climate network, we determine the underlying connectivity as described in Sect.~\ref{sec:methods} for running windows of a given width $w$ in time and study the temporal variability of the resulting global network characteristics.

For comparing the topological properties of evolving climate networks, two different methodological settings are possible: 

\begin{enumerate}[(i)]

\item On the one hand, the global threshold $W^*$ used for edge generation can be kept constant. In this case, we expect variations in the number of edges present in the network as previously found by other authors \cite{Tsonis2008b,Yamasaki2008,Gozolchiani2008,Yamasaki2009} related with the global signature of ENSO variability. We will specifically discuss this situation in Sect.~\ref{sec:fixedthres}. 

\item On the other hand, many complex network characteristics depend on the number of vertices and edges present in the network (cf.~our discussion on the Hamming distance in Sect.~\ref{sec:hamming}). Hence, comparing the properties of climate networks with different numbers of edges and thoroughly interpreting the corresponding results can be a non-trivial task. Therefore, it is desirable to keep the edge density $\rho$ of the networks fixed when studying their time evolution~\footnote{Similar considerations apply, for example, to the case of complex networks obtained from time series of the same system, but with different control parameters, or different parts of the same time series \cite{Donges2011NPG}.}. In this case, the threshold $W^*$ varies in time. A higher threshold thus implies that the empirical distribution $p(W_{\bullet\bullet})$ of the considered pair{-}wise statistical association measure is shifted towards higher values of $W_{\bullet\bullet}$. Thus, periods with increased $W^*$ indicate that there is a higher fraction of strong statistical associations in the climate system, i.e., the obtained edges represent stronger mutual interdependences.

\end{enumerate}

In the following, we study the resulting properties of the global SATA network based on the reanalysis data set (time resolution $\Delta t=1$ day) projected onto an icosahedral grid with $N=10{,}242$ vertices (i.e., $n=5$ refinement steps of the grid construction algorithm described in Sect.~\ref{sec:construction}). For the network evolution, running windows of width $w=1$ year and offset $\Delta w=30$ days are considered. Network connectivity is established based on the lag-zero cross-correlation{s} $C_{ij}({s}=0)$ between {all} pairs $(i,j)$ of records (the alternative case of maximum cross-correlation after allowing for non-zero lags will be discussed in Sect.~\ref{sec:nonzerolags}). Only the $0.5\%$ strongest pair{-}wise associations between time series are considered as edges ($\rho=0.005$). Such \emph{sparse} climate networks have been introduced and partly studied in previous works \cite{Tsonis2004,Tsonis2008b,Donges2009a,Donges2009b,Donges2011,Feng2012}, where $\rho\sim 0.01\dots 0.05$ or $W^*=0.5$ have been typical choices. A brief discussion of climate networks with higher edge densities can be found in Sect.~\ref{sec:extension}~\footnote{We note that there is no generally accepted criterion for network sparsity so far. In the context of growing networks, $\rho\lesssim\mathcal{O}(N^{-1})$ is often considered as sparse, which is supported by the fact that self-organizing networks in various fields show a tendency of an approximate scaling behavior $\rho\sim N^{-1}$~\cite{Laurienti2011}. However, in our case of a fixed network size, these considerations may not apply, so that we distinguish between sparse and dense networks using more heuristic and context-dependent considerations. Notably, given our network size $N$, we would have $\rho\approx 0.0008$, which is about one order of magnitude below the operational range used in our basic setting.}.

In order to guarantee that the climate network at a given point in time only considers dynamical information of its past, we will display the network measures at the \emph{endpoint} of the associated running window.

\subsection{A conceptual view on sparse climate networks}\label{sec:model}

\begin{figure}
\begin{center}
\includegraphics[width=0.485\textwidth]{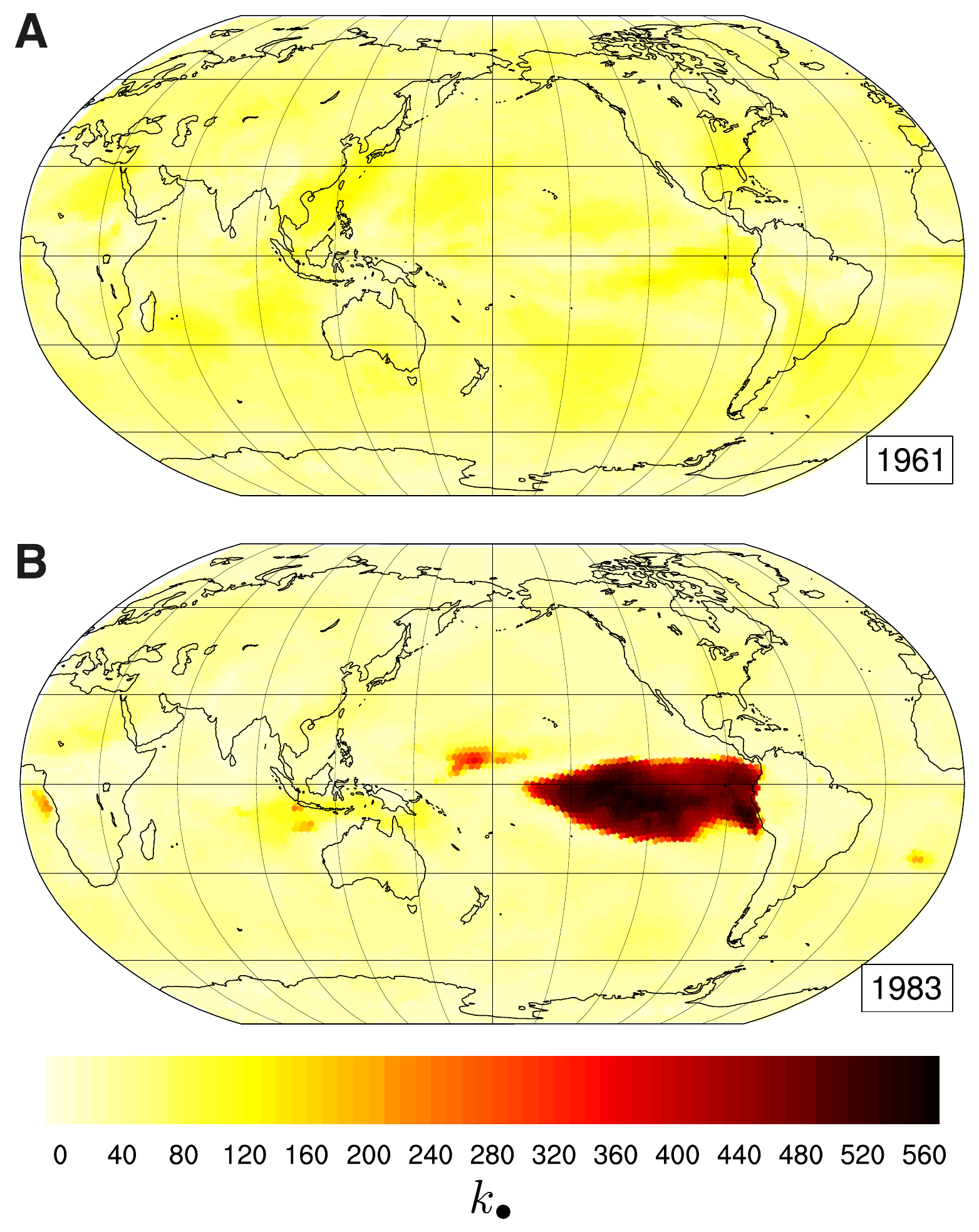}
\end{center}
\caption{(Color online) Spatial distribution of the degree $k_\bullet$ for two typical time windows (A) without (May 1960 to April 1961) and (B) with (May 1982 to April 1983) marked localized structures for the surface air temperature anomalies (SATA) network obtained using the setting described in Sect.~\ref{sec:timdep}.}
\label{fig:fields}
\end{figure}

Before investigating the time-dependence of global climate network characteristics, let us have a detailed look at the spatial patterns associated with the connectivity of these networks. In the following, we will provide a general discussion of these patterns. Thereby, we obtain a conceptual view on sparse climate networks, which will subsequently prove to be helpful for understanding the temporal variability of evolving climate network properties.

When looking at the evolution of spatial connectivity patterns, we find two prototypical phases of the sparse climate network (Fig.~{\ref{fig:fields}}A,B). For certain episodes, it reveals one (or more) distinct strongly connected region(s), i.e., with vertices having extraordinarily high degrees $k_i$ (Fig.~{\ref{fig:fields}}B), while such are not present during other periods (Fig.~{\ref{fig:fields}}A). Inspired by this observation, we propose a simple idealized view on this phenomenon: Certain instances of an evolving climate network -- constructed in the way outlined in Sect.~\ref{sec:timdep} -- exhibit (at least) two types of (temporarily) coexisting structures.

First, there is a ``substrate lattice'' which reflects strong short-range associations between mutually close grid points affected by the same atmospheric circulation patterns. Typically, we observe an approximately exponential decay of the strengths of statistical associations between vertices with increasing distance \cite{Donges2009a}, {since} shorter distances between grid points are typically accompanied by stronger associations between the respective temporal climate variability. Hence, the substrate lattice describes ``trivial'' spatial correlations due to typical (synoptic-scale) atmospheric patterns. We emphasize that this type of structure is always present in our climate networks and behaves relatively static, i.e., its edges do not fluctuate much in time. Further research should clarify the relation to the concept of a ``skeleton of strongly correlated links'' as introduced in \cite{Gozolchiani2008}.

Second, there are regions of larger spatial extension ($\approx 3000-9000$ km){,} which display very high internal connectivity \cite{Tsonis2004,Donges2009a}. The presence of such ``localized structures'' indicates that the spatial correlation length is significantly enhanced within a confined region, i.e., beyond typical synoptic scales. Hence, the corresponding connectivity covers both short (synoptic-scale) and intermediate distances (see Sect.~\ref{sec:timdep:results} for a detailed discussion). Note that localized structures appear only episodically in the {evolving} SATA network (cf.~Fig.~\ref{fig:fields}A,B and \cite{Tsonis2008b}), {but typically repeatedly in the same region (especially the Eastern Equatorial Pacific). As a consequence, we expect them to contribute significantly to the climate network connectivity when considering the full 62 years-long records, which is supported by other studies \cite{Donges2009a,Palus2011}}.

\begin{figure}[t]
\centering
\includegraphics[width=0.485\textwidth]{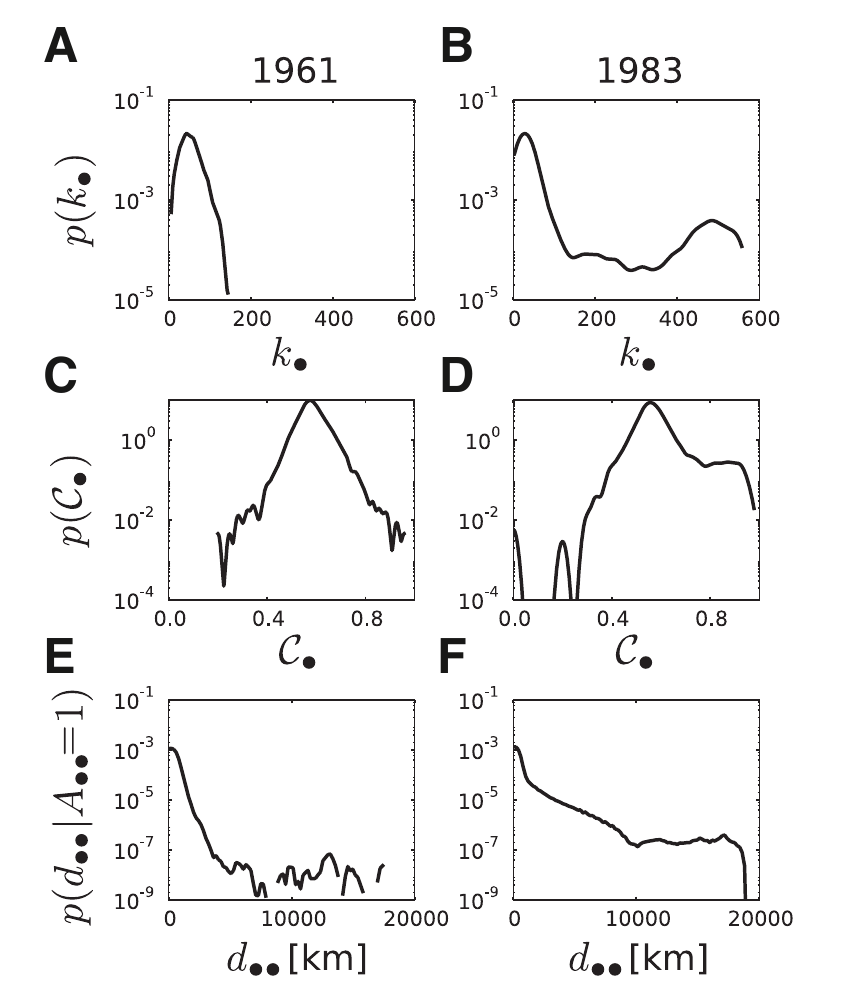}
\caption{Kernel estimates of the probability density functions $p(\cdot)$ (obtained using a Gaussian kernel function with a bandwidth following Scott's rule \cite{Scott2008}) of (A,B) vertex degree $k_\bullet$, (C,D) local clustering coefficient $\mathcal{C}_\bullet$, and (E,F) edge lengths $d_{\bullet\bullet}|A_{\bullet\bullet}=1$ for the same time intervals and setting as in Fig.~\ref{fig:fields}. In all cases, the empirical distributions have been normalized by the distribution of edge lengths of a fully connected graph in order to eliminate purely geometric effects.}
\label{fig:pdfs} 
\end{figure}

The postulated separation of the climate network into substrate lattice and localized structures is supported by the (evolving) edge length distribution (see Fig.~{\ref{fig:pdfs}}E,F), showing one dominant peak for short-range edges (substrate lattice) and far less longer connections. In addition, there are edges of lengths that exceed the typical extension of the described localized structures, {which are} denoted as ``teleconnections'' {and} interrelate climate variability at distant parts of the globe. We hypothesize that the latter show more ``ordered'' placement during certain climatic episodes, although they cannot be clearly separated by means of the edge length distribution only.

Localized structures seem to be favored starting points of long-range edges. For example, the phasing on the El Ni{\~{n}}o Southern Oscillation (ENSO) in the Equatorial Pacific (see Fig.~\ref{fig:fields}B) is known to have considerable influence on climate variability in distant parts of the Earth \cite{Clarke2008}, such as the Indian monsoon system \cite{Maraun2005,Mokhov2010}.

The proposed qualitative view is consistent with previous results for static (time-independent) climate networks, which clearly demonstrated that the majority of vertices is characterized by low connectivity \cite{Tsonis2004,Tsonis2006,Tsonis2008b,Donges2009a} (see also Fig.~{\ref{fig:pdfs}}A,B). Furthermore, the localized structures in the Eastern Equatorial Pacific (Fig.~{\ref{fig:fields}}B) related with ENSO variability closely resemble the corresponding results of recent studies \cite{Tsonis2004,Donges2009a}. Notably, these observations hold for our analysis using an icosahedral grid, whereas former studies were based on on a standard (angularly regular) grid.

\begin{figure}[t]
\centering
\includegraphics[width=0.485\textwidth]{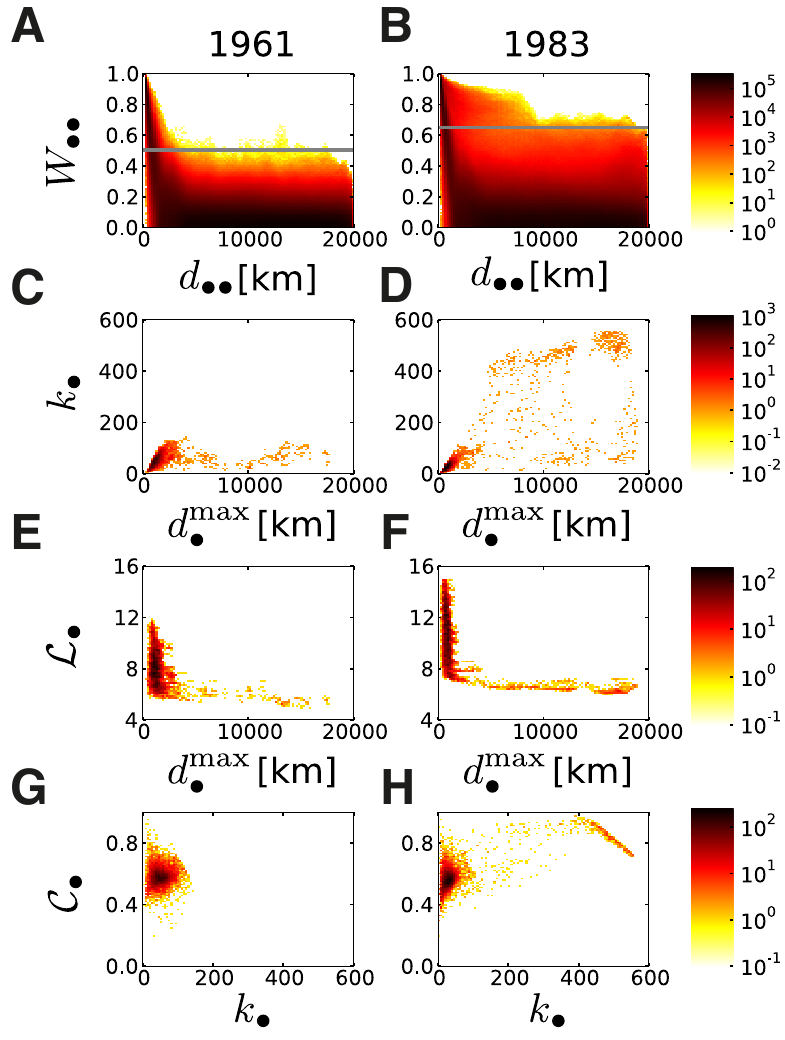}
\caption{(Color online) Joint probability distributions $p(\cdot, \cdot)$ of (A,B) strength of statistical association $W_{\bullet\bullet}$ and geographical distance $d_{\bullet\bullet}$ between all pairs of vertices (i.e., without thresholding), and (C,D) degree $k_\bullet$ and maximum edge length $d_\bullet^{\text{max}}$, (E,F) average shortest-path length per vertex $\mathcal{L}_\bullet$ and maximum edge length $d_\bullet^{\text{max}}$, and (G,H) local clustering coefficient $\mathcal{C}_\bullet$ and degree $k_\bullet$ of all vertices (i.e., after thresholding), for the same time intervals and setting as in Fig.~\ref{fig:fields}. All distributions are represented as histograms using 100 (90 for $W_{\bullet\bullet}$) equidistant bins. The grey line{s} in (A,B) {depict} the threshold{s} above which edges are established (here: $0.5\%$ of all possible pairs of vertices). Wherever appropriate, the distance-dependences of the distributions have been corrected as in Fig.~\ref{fig:pdfs}.}
\label{fig:degree_clustering} 
\end{figure}

Notably, our conceptual view refers to the membership of vertices to one or another category, but is induced by the placement and temporal behavior of edges. In turn, analyzing {fields of} (topological or geographical) vertex properties does only provide implicit information on the edges. However, even though the degree field (Fig.~{\ref{fig:fields}}) does not describe the spatial distribution of long-range connections (Fig.~{\ref{fig:pdfs}}E,F), our idealized conceptual view holds, since the joint distribution of maximal edge length per vertex $d_\bullet^{\text{max}}$ and degree $k_\bullet$ (Fig.~\ref{fig:degree_clustering}C,D) shows that the vast majority of vertices with small degree has indeed almost no long-range edges ($d_i^{\text{max}} \leq 2500$ km). Further relationships between network properties will be discussed below.

\subsection{Temporal variability of global network properties} \label{sec:timdep:results}

\begin{figure}[t]
\centering
\includegraphics[width=0.485\textwidth]{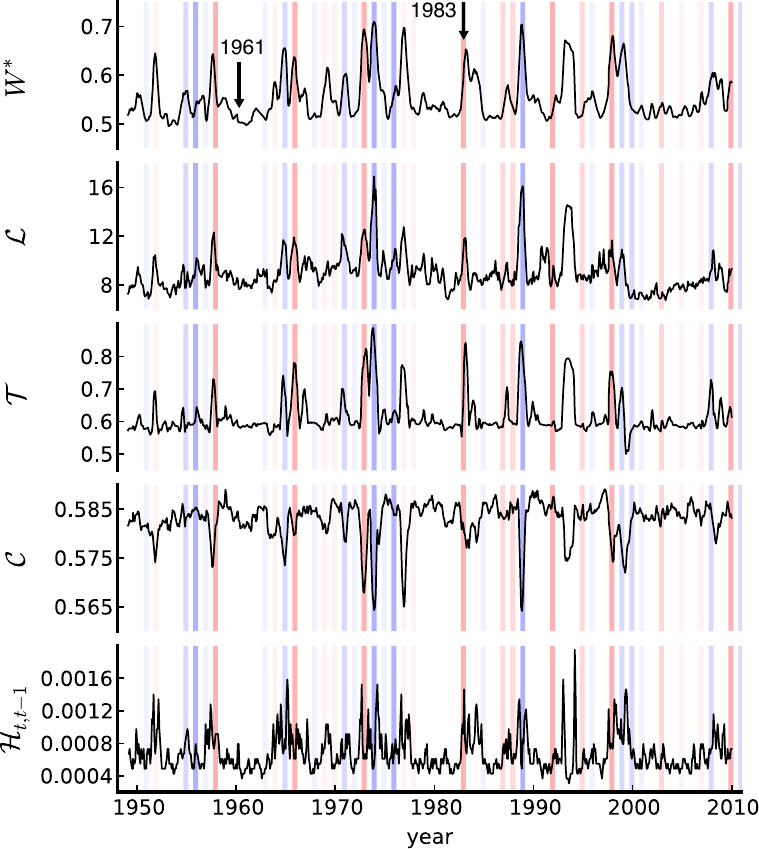}
\caption{(Color online) Threshold $W^*$, average path length $\mathcal{L}$, transitivity $\mathcal{T}$, global clustering coefficient $\mathcal{C}$ and (standard) Hamming distance $\mathcal{H}_{t,t-1}$ between networks obtained for successive periods in time for the evolving SATA network (using {the} settings given in Sec.~\ref{sec:timdep}).
Note that due to the fixed $\rho$, we have $\mathcal{H}_{t,t-1}^*=\mathcal{H}_{t,t-1}$. Vertical bars indicate fall and winter seasons (SON-DJF) with maximal intensity of EN (red) and LN (blue; darker colors represent stronger episodes according to the Ni\~no 3.4 index, cf. \cite{Trenberth1997,Rayner2003}).}
\label{fig:evolution1} 
\end{figure}

Performing an evolving climate network analysis as described above, we first observe that the two network measures $\mathcal{L}$ and $\mathcal{T}$ as well as the Hamming distance $\mathcal{H}_{t,t-1}$ widely change in parallel with each other and with the threshold $W^*$, with characteristic peaks from a certain constant base level  (Fig.~\ref{fig:evolution1}). We emphasize that this co-evolution is \emph{ex ante} non-trivial, since these three measures capture distinctively different network properties. In turn, the variability of the global clustering coefficient $\mathcal{C}$ is strongly anti-correlated with that of the aforementioned characteristics, which also deserves further discussion since $\mathcal{C}$ captures a similar network property as $\mathcal{T}$. 

Because the total number of edges has been kept fixed, all scalar network characteristics are not affected if the edge density $\rho$ is varied within a certain range still corresponding to a ``sparse'' connectivity. Hence, the strong similarity between the variations of both $\mathcal{L}$ and $\mathcal{T}$ in the climate network does most probably originate from complex rewiring processes driven by climate variability, although we cannot fully rule out minor effects due to changing auto-correlations. In the following, we will provide a detailed graph-theoretical interpretation of these results, whereas the climatological mechanisms beyond the obtained temporal variability pattern will be discussed in detail in Sect.~\ref{sec:discussion}.

\subsubsection{Association strengths and spatial scales}

Since the SATA networks studied in this work solely rely on those pairs of time series the statistical association between which exceeds $W^*$, the evolving joint probability density function $p(d_{\bullet\bullet}, W_{\bullet\bullet})$ (Fig.~\ref{fig:degree_clustering}A,B) reveals first deep insights into relevant spatio-temporal modes of climate variability. Specifically, this distribution can be qualitatively decomposed into the components introduced in Sec.~\ref{sec:model}: The substrate lattice manifests itself as dominant strong and rather persistent associations at small edge lengths. For larger edge lenghts, there is a more or less continuous distribution of association values. During some periods (e.g., in Fig.~\ref{fig:degree_clustering}B), the corresponding distribution of statistical association values for distant vertices is shifted towards higher values, indicating the presence of localized structures.

\begin{figure}
\centering
\includegraphics[width=0.485\textwidth]{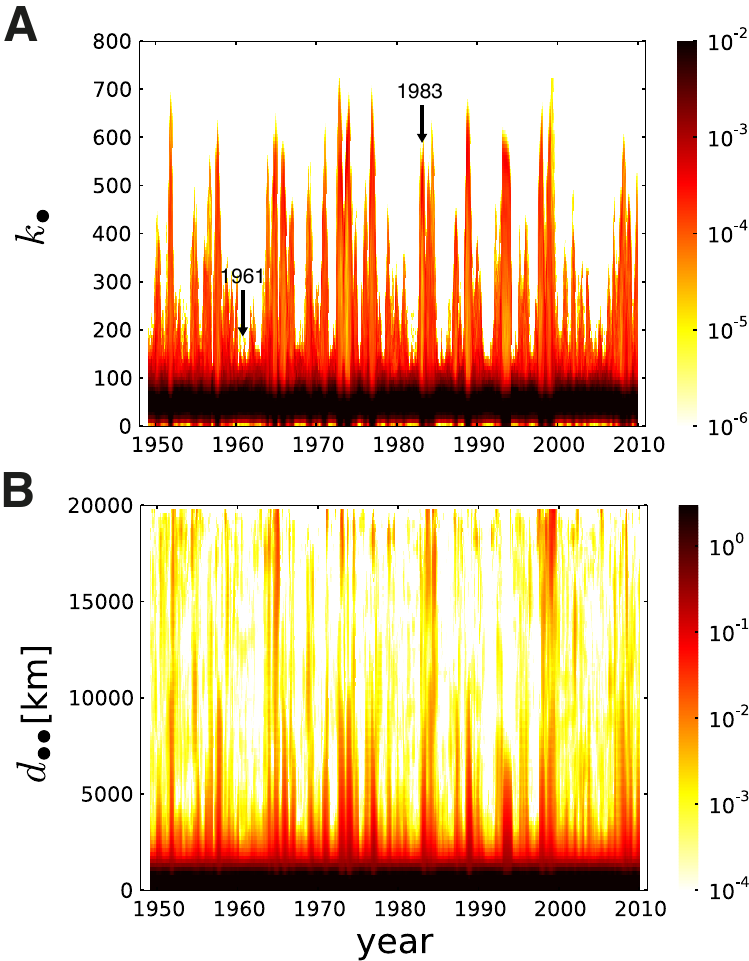}
\caption{(Color online) Evolution of (A) degree distribution $p(k_\bullet)$ and (B) edge length distribution $p(d_{\bullet\bullet}|A_{\bullet\bullet}=1)$ with the settings given in Sect.~\ref{sec:timdep}. In panel (B), the distributions obtained for all time windows have been corrected as in Fig.~\ref{fig:pdfs}.}
\label{fig:distribution-evolving2} 
\end{figure}

Considering the evolution of $p(d_{\bullet\bullet}, W_{\bullet\bullet})$, we see as a first approximation a net amplification of association values for several time windows, leading to the peaks in the threshold $W^*$ visible in Fig.~\ref{fig:evolution1}. This amplification is not uniform with respect to the spatial scale $d_{\bullet\bullet}$. Consequently, not only the degree distribution $p(k_\bullet)$ (Fig.~\ref{fig:distribution-evolving2}A), but also the edge length distribution varies substantially with time (cf. Fig.~\ref{fig:distribution-evolving2}B). For several time windows exhibiting strong peaks in the evolving scalar network characteristics, we observe more long edges -- but yet cannot find a clear separation of the longer spatial scales. This suggests that these time windows are accompanied by the emergence of localized structures and, hence, hub vertices (Fig.~\ref{fig:distribution-evolving2}A). Recall that localized structures consist of vertices with very high degrees (cf. Fig.~\ref{fig:pdfs}B) and exhibit high internal connectivity. Typically, but not necessarily, the associated structures are located in the {E}quatorial Pacific (cf. Fig.~\ref{fig:fields}B).

\subsubsection{Transitivity}

By forming groups of vertices with very high degree (commonly in the presence of localized structures), the network`s total number of connected triples rises, since the possible number of triples centered at one particular vertex $i$ grows with its degree $k_i$ as $k_i(k_i-1)/2$. As a consequence, although the denominator in Eq.~(\ref{eq:transitivity}) peaks at those time windows within which the strength of statistical associations is amplified (peaking $W^\star$), the transitivity $\mathcal{T}$ still increases because the total number of closed triangles (numerator in Eq.~(\ref{eq:transitivity})) increases even stronger than the number of connected triples.

In analogy with the local clustering coefficient (Eq.~(\ref{eq:clustering})), we can formally split the transitivity (Eq.~(\ref{eq:transitivity})) into (non-normalized) ``local'' transitivities $\mathcal{T}_i$ by just decomposing the sum in the nominator of Eq.~(\ref{eq:transitivity}) as
\begin{equation}
\mathcal{T}=\frac{1}{N}\sum_{i=1}^N \mathcal{T}_i \quad \mbox{with} \quad
\mathcal{T}_i=\frac{\sum_{j,k=1}^N A_{ij} A_{ik} A_{jk}}{\frac{1}{N}\sum_{i,j,k=1, j \neq k}^N A_{ij} A_{ik}},
\label{eq:localtransitivity}
\end{equation}
\noindent
i.e., $\mathcal{T}_i$ gives the ratio between the number of triangles centered at vertex $i$ and the average number of connected triples centered at all vertices. We find that vertices $i$ with high degree in the localized structures contribute stronger (i.e., with higher $\mathcal{T}_i$) to the overall transitivity than those exclusively belonging to the substrate lattice.

\subsubsection{Global clustering coefficient}

Unlike network transitivity, the global clustering clustering coefficient $\mathcal{C}$ -- as the arithmetic mean of all local clustering coefficients -- drops when localized structures emerge. This behavior appears somewhat unexpected, since both characteristics quantify conceptually related properties and exhibit values within the interval $[0,1]$. In connection with this fact, note that the variability of $\mathcal{C}$ is by about one order of magnitude smaller than that of $\mathcal{T}$, another observation that calls for explanation.

In order to resolve the reason for the behavior described above, a deeper look into the probability distribution of $\mathcal{C}_\bullet$ (Fig.~\ref{fig:pdfs}C,D) gives a two-fold finding: For several time windows we observe a secondary maximum of $p(\mathcal{C}_\bullet)$ at higher $\mathcal{C}_\bullet$ as well as a shift of the primary maximum towards smaller values. If we furthermore consider the dependence on the vertices' degrees $k_\bullet$, we find that the hubs show a broad range of higher $\mathcal{C}_\bullet$ values than the vertices exclusively belonging to the substrate lattice (cf. Fig.~\ref{fig:degree_clustering}H). Still, the vast majority of vertices with low degrees show declining $\mathcal{C}_\bullet$ during periods with marked localized structures. This causes the global clustering coefficient to drop (even though a notable fraction of vertices increase their $\mathcal{C}_\bullet$). We can exclude that the observed drops in $\mathcal{C}$ have been induced by vertices $i$ of degree $k_i=0$ or $k_i=1$, since only five of such vertices emerge in the entire time-evolution of the SATA network. 

According to the general shape of the probability density $p(d_{\bullet\bullet},W_{\bullet\bullet})$ (Fig.~\ref{fig:degree_clustering}A,B), for a \emph{fixed} time window we can expect that the distributions of $k_\bullet$ and $\mathcal{C}_\bullet$ will exhibit remarkable changes as the edge density $\rho$ is varied. As a consequence, we hypothesize {that} the general behavior of $\mathcal{C}$ (for fixed $\rho$ as a function of time) is much more strongly affected by the specific choice of $\rho$ than that of $\mathcal{T}$. The validity of this hypothesis, and particularly the dependence of the distinct anti-correlation between $\mathcal{T}$ and $\mathcal{C}$ on the chosen edge density, will be further discussed in Sec.~\ref{sec:extension}.

\subsubsection{Average path length}

Ad hoc it seems counter-intuitive that a spatially embedded network (with periodic boundary conditions and fixed edge density) exhibits a \emph{rising} topological path length when there is a transfer of connectivity towards \emph{longer} spatial scales. Specifically, in spatially embedded networks (e.g., airline transportation), longer edges typically act as shortcuts. Thus, the presence of such long-range connections is particularly reflected in shortest path-based quantities. In our SATA networks, the same observation holds in each time window: vertices $i$ with $d_i^{\text{max}} \gtrsim 2500$ km have always minimal $\mathcal{L}_i$. However, at the same time we observe a total shift of $p(d_\bullet^{\text{max}}, \mathcal{L}_\bullet)$ towards higher $\mathcal{L}_\bullet$ values in these time windows (Fig.~\ref{fig:degree_clustering}E,F). We deduce, that this can be caused by a more redundant, partially parallel geographical placement of long-range edges compared to the base-level situation (cf. Fig.~\ref{fig:fields}). This explanation is consistent with the physical continuity of the climate system: Spatially close points tend to behave similar and thus correlate group-wise with others. Another possible cause is the loss of edge density in the substrate lattice, enlarging shortest paths starting or ending at (the majority of) vertices $i$ with small $k_i$. In a nutshell, building up lots of parallel highways by dismantling rural roads does stretch shortest pathways in the entire frame.

\subsubsection{Hamming distance}

In a similar spirit as for the global network characteristics discussed above, the peaks of the Hamming distance $\mathcal{H}_{t,t-1}$ (Eq.~(\ref{def:hamming})) coinciding with those of $W^*$ can be explained as indicators of a persistent redistribution of edges between different spatial scales, which is known to be a typical signature of EN episodes \cite{Tsonis2008b,Yamasaki2008,Gozolchiani2008}. Specifically, $\mathcal{H}_{t,t-1}$ exhibits a double-peak structure around time windows characterized by single peaks of the other network characteristics ($\mathcal{T}$, $\mathcal{C}$ and $\mathcal{L}$). Given our interpretation of the Hamming distance as a ``network derivative'', this finding is consistent with the expected behavior: large values coincide with time periods where the SATA network connectivity is changing considerably, i.e., in parallel with the emergence and disappearance of localized structures.

\subsection{Possible implications for local network organization}

Following the observations described in Sect.~\ref{sec:timdep:results} in combination with our conceptual view on sparse climate networks (Sect.~\ref{sec:model}), we are able to derive some preliminary insights into the spatial organization of the association structure of SATA fields on the local (network) scale, which complement recent findings \cite{Donges2009b}. For this purpose, let us examine Fig.~\ref{fig:degree_clustering} in some more detail.

In many examples of complex networks \cite{Ravasz2002,Ravasz2003,Dorogovstev2002,Szabo2003,Vazquez2003}, hubs have a tendency to contribute to a lower fraction of triangles than vertices with intermediate degree. In the climate network, this effect is only visible for those vertices $i$ with the highest degrees (i.e., $k_i\gtrsim 450$ in Fig.~\ref{fig:degree_clustering}H), which belong to densely connected and spatially localized structures (see Fig.~\ref{fig:fields}B). As discussed in Sect.~\ref{sec:timdep:results}, these hubs have higher local clustering coefficients than the substrate lattice, even though the associated spatial scales captured by the adjacent edges are considerably larger than the typical ``correlation range'' (i.e., synoptic scales of up to $\mathcal{O}(10^3$ km)) within which mutual associations are \textit{on average} statistically relevant. Notably, this effect acts against the decrease of the local clustering coefficients in the substrate lattice, which dominates the resulting signature in the global clustering coefficient for sparse climate networks with an edge density of 0.5\% as considered here. 

In order to derive an alternative explanation, note again that in the presence of localized structures, vertices belonging to the substrate lattice are characterized by a lower average degree than otherwise, since the total edge density $\rho$ is conserved (compare the left and right panels in Figs.~\ref{fig:pdfs} and \ref{fig:degree_clustering}). We suggest that this finding could indicate that the connectivity in the substrate lattice becomes less isotropic, but rather reflects the actual prefer{r}ed directions of atmospheric dynamics (e.g., westerlies, trade winds, etc.). This hypothesis is supported by recent findings of Palu\v{s} \textit{et~al.}~\cite{Palus2011} who, by using a different climate network approach, observed that a stronger transport of air masses during positive phases of the North Atlantic Oscillation (NAO) enhances the network connectivity in the directly affected areas. Particularly, it is likely that vertices aligned in parallel with the prefer{r}ed direction of atmospheric flow have on average stronger statistical associations over a wider spatial range than those in the perpendicular direction. A detailed examination of the associated climate network connectivity patterns on the local scale is, however, beyond the scope of this study and will be subject of future work. We conjecture that in addition to established vertex characteristics, purely geometric measures related to the spatial anisotropy of connections \cite{Chan2011,Kutza2012} can provide relevant complementary information for this purpose.

\subsection{Statistical interdependences between measures}

\begin{figure}
\centering
\includegraphics[width=0.485\textwidth]{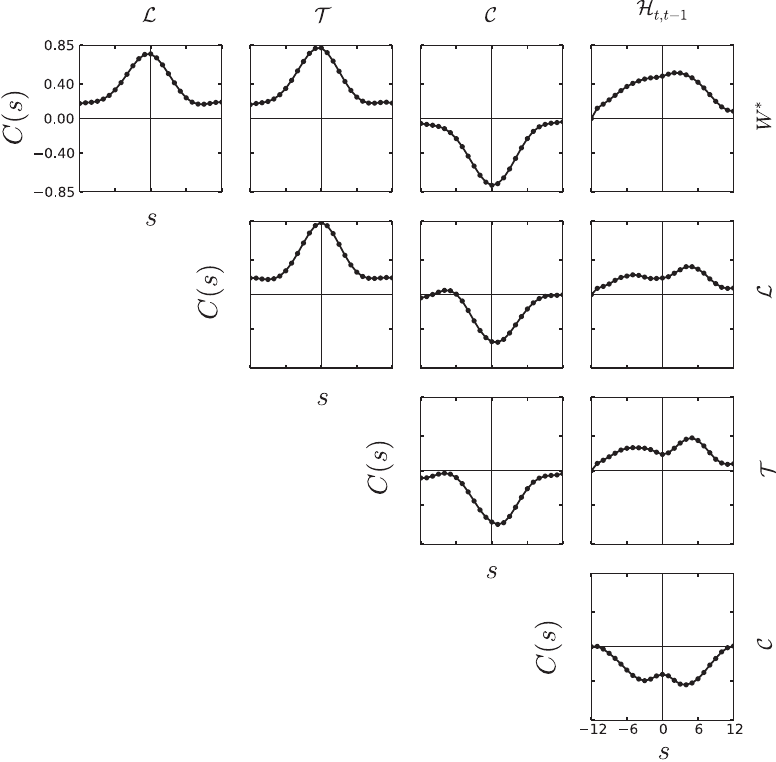}
\caption{Linear cross-correlation functions $C({s})$ between the different evolving climate network characteristics from Fig.~\ref{fig:evolution1}. Positive (negative) ${s}$ refer to the situation of the measure in the row leading (lagging) the measure in the column.}
\label{fig:correlations1} 
\end{figure}

Beyond the qualitative interpretation of the observed similarities and dissimilarities of various characteristics of evolving climate networks, we will next provide a quantitative assessment of the statistical interrelationships between these measures{,} empirically supplementing our arguments from the former sections. For this purpose, Fig.~\ref{fig:correlations1} displays the linear cross-correlation functions between the {temporal variability of} different measures. For the global network characteristics{,} the most significant (positive or negative) interdependences are found when considering the same time window. In contrast, for the Hamming distance, the maximum correlations with the other considered measures show a delay between $3 \Delta w$ and $6 \Delta w$ (i.e., of 3-6 months), underlining the distinctively different meaning of $\mathcal{H}_{t,t-1}$ as a ``network derivative'' indicating structural changes before and after their most significant reflection in the global network characteristics. In this respect, the observed delay reflects the typical time-scale associated with the emergence and disappearance of localized structures and, thus, of the underlying climate phenomena (cf.~Sect.~\ref{sec:discussion}).

\section{Robustness of the results}\label{sec:robustness}

The results described so far have been obtained using one specific setting of methodological options for climate network construction. In the following, we test the qualitative robustness of the obtained results using different methodological choices.

\subsection{Non-zero lags}\label{sec:nonzerolags}

\begin{figure}
\centering
\includegraphics[width=0.485\textwidth]{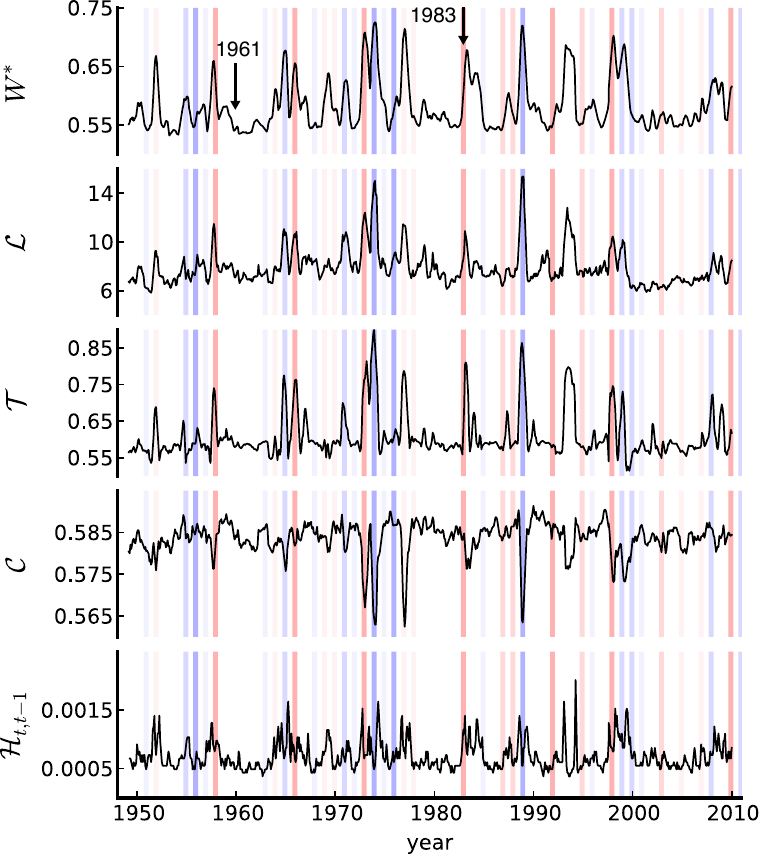}
\caption{(Color online) As in Fig.~\ref{fig:evolution1} for SATA networks based on the maximum cross-correlation for time lags ${s} \in [0, 30]$ days. Note that $W^*$ is larger than in Fig.~\ref{fig:evolution1} for ${s}=0$ as expected.}
\label{fig:evolution2} 
\end{figure}

Our results in Sect.~\ref{sec:results} refer to climate networks based on lag-zero statistical associations between first-order deseasoned SAT time series at distinct parts on the globe. Since atmospheric circulation patterns always travel with a finite velocity, the same variations usually affect different grid points at different times, so that the cross-correlation function $C_{ij}({s})$ between different grid points $i$,$j$ could peak at non-zero mutual lags (${s}\neq 0$). In order to study the impact of such lags on the topology and time evolution of the SATA network, in the following we replace the lag-zero cross-correlation $C_{ij}({s}=0)$ as the criterion for edge creation by the maximum value of the cross-correlation function $C_{ij}({s})$ for time lags ${s}\leq 30$ days. This choice allows considering typical large-scale atmospheric wave phenomena that could mediate between the temperature variability at distant parts on the Earth, and respects the typical lifetime of weather regimes. Keeping all other parameters of our analysis the same as above, the resulting variations of climate network properties are shown in Fig.~\ref{fig:evolution2}. 

As for the lag-zero case, we observe a sharp increase of the threshold $W^*$ at the previously identified time windows, which mainly coincide with certain phases of the El Ni{\~{n}}o Southern Oscillation (ENSO, see Sec{t}.~\ref{sec:discussion}). {Related to this finding, we note that other authors (e.g.,~\cite{Yamasaki2008,Gozolchiani2008}) even} found signatures corresponding to a \emph{decrease} rather than increase in $W^*$. 

{On the one hand, Tsonis and Swanson \cite{Tsonis2008b} as well as Palus \textit{et~al.}~\cite{Palus2011} compared climate networks obtained for EN and LN phases and found higher connectivity during LN in comparison with EN episodes. For a fixed edge density, this would correspond to higher values of $W^*$. However, the aforementioned studies did not explicitly consider the neutral state as opposed to both EN and LN conditions. In turn, the results of our evolving network analysis displayed in Figs.~\ref{fig:evolution1} and \ref{fig:evolution2} do not allow a systematic confirmation or rejection of any particular asymmetry between the values of $W^*$ or the considered network characteristics for all EN and LN phases, although there are two LN episodes (1973/74 and 1988/89) that exhibit higher $W^*$ values than during all EN phases. We note that these results are consistent with recent findings of Martin \textit{et~al.}~\cite{Martin2013}, who observed higher correlations during EN periods than under ``normal'' climate conditions, and even stronger pair-wise associations during LN phases in agreement with \cite{Tsonis2008b,Palus2011}.} 

{On the other hand,} \cite{Yamasaki2008,Gozolchiani2008} considered a setting with {a} fixed threshold rather than {a} fixed edge density. Since $W^*$ and $\rho$ are closely interrelated, {the} decrease in the number of edges during EN phases {reported in these studies would coincide} with a decrease in $W^*$ if $\rho$ is kept constant, {which is different from our results. One possible reason for this difference} is that we prefer \textit{not} to normalize the estimated association strengths $W_{ij}$ by the standard deviations of the measure taken over all considered time lags {as in \cite{Yamasaki2008,Gozolchiani2008}. Specifically,} we can argue that the absolute value of the maximum statistical association and its magnitude relative to the fluctuations over a range of delays (as defined in \cite{Yamasaki2008,Gozolchiani2008}) provide complementary results. A decreasing relative magnitude in parallel with an increasing absolute value indicates stronger statistical associations for most other delays. Here, we keep the two quantities (i.e., absolute value of maximal statistical association and standard deviation of associations over a certain range of ${s}$) separated.
{This point of view is supported by \cite{Martin2013}, who found that the temporal fluctuations in the network connectivity obtained using the approach of \cite{Yamasaki2008,Gozolchiani2008} do not necessarily reflect changes in the coupling between different regions. Even more, this method appears to have a lower degree of robustness under changes of its basic parameters than other approaches for climate network construction \cite{Martin2013}.}

{For the temporal variation of the considered network characteristics}, we find no qualitative deviations from the findings previously obtained for the lag-zero SATA network (compare Figs.~\ref{fig:evolution1} and \ref{fig:evolution2}). A detailed inspection of the delays associated with the maximum cross-correlation (not shown) or, alternatively, the maximum mutual information \cite{Radebach2010} reveals that besides an exceptionally strong peak around ${s}=0$, almost all delays contribute with comparable frequencies. This observation demonstrates that statistically relevant atmospheric interactions appear predominantly on very short time scales, reflecting the presence of particularly strong interactions between geographically close grid points, and subsequently on all other (here considered) time scales. By this, we qualitatively reproduce \cite{Yamasaki2008,Gozolchiani2008}. Note that the absence of marked changes on the global network scale in comparison with the lag-zero case does not necessarily imply that there are no changes at the local scale. A deeper discussion of the associated fields of local (vertex) characteristics is to be resumed in future work.

\subsection{Fixed thresholds}\label{sec:fixedthres}

\begin{figure}
\centering
\includegraphics[width=0.485\textwidth]{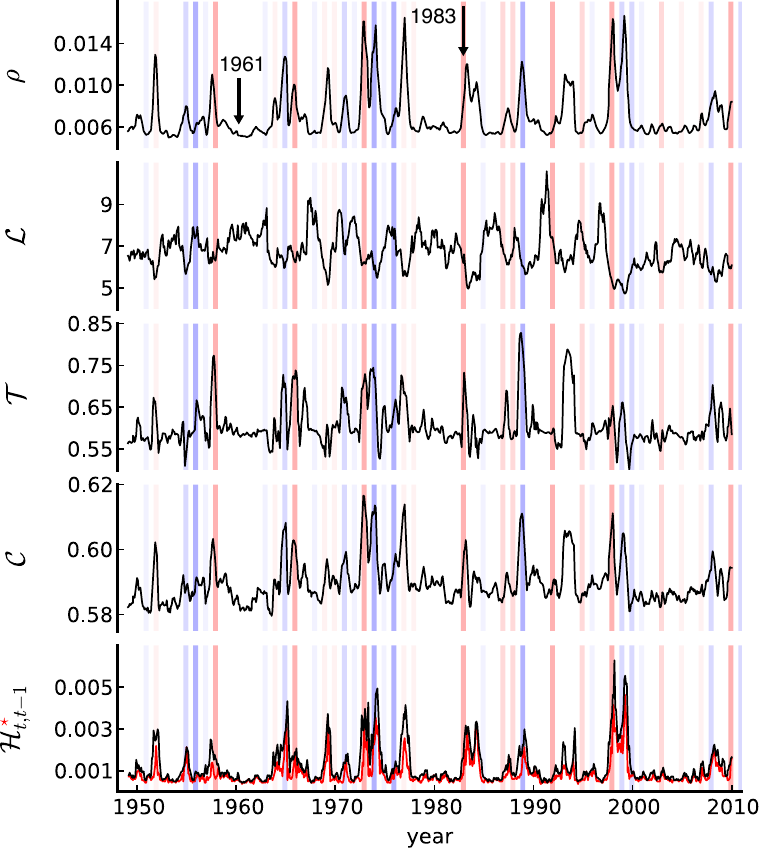}
\caption{(Color online) As in Fig.~\ref{fig:evolution1} for SATA networks with a fixed threshold $W^*$ chosen as the minimal threshold from the evolution for fixed edge density of $\rho=0.005$ (see topmost panel in Fig.~\ref{fig:evolution1}), $W^*\approx 0.4958$ -- a value which was also used in \cite{Tsonis2004,Tsonis2006,Tsonis2008b}. Thus, $\rho=0.005$ is a lower boundary for the evolving edge density, ensuring that the network does not tend towards disintegration. In addition to the standard Hamming distance $\mathcal{H}_{t,t-1}$ (lower panel, black line), the corrected Hamming distance $H^*_{t,t-1}$ is shown (red line).}
\label{fig:evolution3} 
\end{figure}

As initially discussed in Sect.~\ref{sec:timdep}, there are two possible and theoretically justified options for selecting a global threshold $W^*$ in evolving climate network analysis. While all previous considerations have focused on a fixed edge density $\rho$ and, hence, a variable threshold $W^*$, in the following we consider the alternative choice of a fixed threshold $W^*$, which in turn implies that $\rho$ becomes time-dependent. We emphasize that the resulting variations of $W^*$ and $\rho$, respectively, are directly interrelated, since maxima of both $W^*$ (for fixed $\rho$) and $\rho$ (for fixed $W^*$) indicate a shift of the distribution $p(W)$ of association strengths towards larger values. Consequently, the temporal variability pattern of $\rho$ (see Fig.~\ref{fig:evolution3}) infer{r}ed from lag-zero-based cross-correlation $C_{ij}({s}=0)$ is similar to that of $W^*$ in Fig.~\ref{fig:evolution1}.

A detailed inspection of the different global network characteristics (cf. Fig.~\ref{fig:evolution3}) shows that the behavior of the transitivity $\mathcal{T}$ is qualitatively similar to the case of fixed edge density $\rho$, whereas the global clustering $\mathcal{C}$ exhibits peaks instead of drops in the previously identified time windows, and the average shortest path length $\mathcal{L}$ lacks the formerly observed peaks from a constant base level.

Clearly, the occur{r}ence of localized structures as described in Sec{t}.~\ref{sec:model} takes place in the case of fixed threshold as well. Specifically, in the presence of such structures, we observe a considerably higher edge density. Hence, edges are not only spatially redistributed, but \textit{additional} significant associations emerge.

Since the transitivity $\mathcal{T}$ shows the same signal as before, our considerations from Sec.~\ref{sec:timdep:results} still apply. Specifically, the emergence of localized structures results in a marked increase in local transitivities (even overshadowing a potential decrease of local transitivity at the majority of lower-degree vertices). Thus, additional edges (of lower association strengths) follow the formerly discussed mechanisms.

The switch in the qualitative behavior of the global clustering coefficient $\mathcal{C}$ (from drops to peaks) is due to the fact that the additional edges (which preferentially connect vertices within the localized structures) do not substantially affect the edge structure of the substrate lattice. Consequently, we do \emph{not} find a shift of the primary maximum of $p(k_\bullet, \mathcal{C}_\bullet)$ towards smaller values of $\mathcal{C}_\bullet$, while the secondary maximum shows the formerly described behavior contributing to an overall increase in $\mathcal{C}$.

The average shortest path length $\mathcal{L}$ is governed by two competing mechanisms. While the effect pointed out in Sec.~\ref{sec:timdep:results} (i.e., a spatial redistribution of edges leading to a more redundant placement, yielding an overall increasing path length) is still present, additionally occurring edges trivially reduce the lengths of shortest paths. A further detailed investigation of the distribution of shortest path lengths per vertex could separate the two effect but exceeds the scope of this work.

Since the Hamming distance $\mathcal{H}_{t,t-1}$ as well as its density-corrected counter-part $\mathcal{H}_{t,t-1}^*$ indicate general connectivity changes, both have high values whenever localized structures emerge or disappear. {In general, the amplitudes of $\mathcal{H}_{t,t-1}$ are about a factor of $3$ larger than in the case of fixed edge densities (Figs.~\ref{fig:evolution1} and \ref{fig:evolution2}). This observation is probably related to the fact that  the edge densities in the considered fixed threshold scenario are bound from below by the value used in the fixed edge density scenario. Hence, for most time intervals, there are considerably more edges contained in the networks of Fig.~\ref{fig:evolution3} than in those of Figs.~\ref{fig:evolution1} and and \ref{fig:evolution2} (up to about three times more). Since the higher the edge density, the more edges can be rewired between two subsequent time intervals (note that the range of possible Hamming distances is bound from above by the sum of the edge densities of the two networks to be compared if the latter is smaller than 1), the difference in the realized edge densities could explain the observed behavior.}

{Comparing the classical and corrected Hamming distance, we} find that the contribution of edge rewiring mostly dominates the effect of a changing edge density $\rho$, i.e., both $\mathcal{H}_{t,t-1}$ and $\mathcal{H}_{t,t-1}^*$ display qualitatively the same variability. However, in some time windows (e.g., in 1957, 1976, or 1993), the corrected version of this measure remains at a much lower level (cf.~Fig.~\ref{fig:evolution3}), indicating that during these periods, the changes in the edge density $\rho$ are particularly relevant as well.

\subsection{Further methodological options}

Besides the methodological choices discussed above, there are further options that can be used for modifying the setting of our evolving climate network analysis. In the following, we just briefly note some of the possibilities that have been tested within the course of the described work (see \cite{Radebach2010} for examples), but are not discussed here in detail since they lead to results that are qualitatively equivalent to those already presented above:
\begin{itemize}
\item use another statistical association measure, e.g., the nonlinear (cross-) mutual information or Spearman correlation coefficient (functions) instead of linear Pearson correlation, further measures are possible (cf.~Sect.~\ref{sec:edges}),
\item change the temporal resolution of the considered time series data (e.g., 6 hours or one month),
\item change the size $w$ of the running windows in time used for the evolving network analysis within a reasonable range.
\end{itemize}
\noindent
In contrast to the aforementioned options, the choice of the spatial resolution of the icosahedral grid (e.g., use $n=4$ or $n=6$ instead of $n=5$ refinement steps, cf.~Sect.~{\ref{sec:construction}}) is crucial. Using a coarser grid than presented here leads to a significantly increased fraction of (almost) disconnected nodes ($k_i=0$ or $k_i=1$) for the edge densities used in this study, especially in the presence of localized structures. This effect then biases other network measures (e.g., the global clustering coefficient $\mathcal{C}$) and reduces accessible information about the network's structure. In turn, considering a finer grid would require data provided with a higher spatial resolution. For the large-scale global climate characteristics we are interested in, the considered resolution is reasonable{,} whereas consideration of specific regional atmospheric processes and associated statistical association patterns \cite{Malik2010,Malik2011,Boers2013,Rheinwalt2012} would call for a denser grid.


\subsection{Denser climate networks}\label{sec:extension}

All previous considerations referred to sparse evolving climate networks, i.e., networks with a very low edge density. While there have been several studies using this setting (e.g.,~\cite{Tsonis2004,Tsonis2008b,Donges2009a,Donges2009b,Donges2011,Feng2012}), one might argue for analyzing denser networks (as used in~\cite{Carpi2012EPJB,Rheinwalt2012}) in order to obtain possibly better statistical estimates of network characteristics. In the following, we discuss to which extent our results described above are modified in case of higher edge densities.

\begin{figure}
\centering
\includegraphics[width=0.485\textwidth]{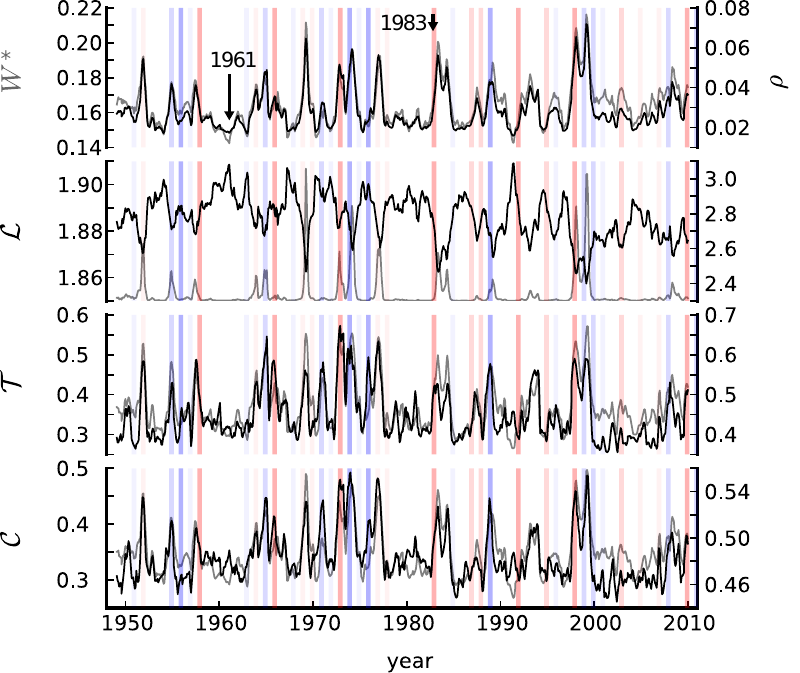}
\caption{(Color online) Global network measures (without Hamming distance) for denser climate networks. The displayed results correspond to a fixed edge density $\rho=0.15$ (left axes, gray lines) and a fixed threshold $W^*=0.2997$ (right axes, black lines), repectively.}
\label{fig:extension} 
\end{figure}
 
As Fig.~\ref{fig:extension} shows, we find that many of the previously discussed signatures of global network characteristics are qualitatively robust when considering higher fixed edge densities up to $\rho=0.25$, as well as lower fixed thresholds (e.g., $W^*=0.2997$, corresponding to $\rho\in(0.016,0.072)$). This observation partially confirms the previous result of \cite{Tsonis2010} that ``the effect of different correlation thresholds (between 0.4 and 0.6) does not affect the conclusions reached''. Specifically, we make the following observations:

\begin{enumerate}[(i)]

\item In accordance with its previously discussed robustness against different methodological choices (e.g., in Sec.~\ref{sec:fixedthres}), we find the transitivity $\mathcal{T}$ to be the most robust measure. Not only for relatively large edge densities, but also at extremely high thresholds ($W^*=0.9$, corresponding to $\rho\in(0.0004,0.0010)$), the evolution of Fig.~\ref{fig:evolution1} is confirmed. 

\item In contrast to this, the global clustering coefficient $\mathcal{C}$ shows a marked sensitivity to variations of the edge density: it drops during certain time intervals for low edge densities (e.g., $\rho \lesssim 0.01$, cf.~Sec.~\ref{sec:timdep:results}), but peaks for higher ones. The critical edge density at which this behavior switches is determined by $p(d_{\bullet\bullet}, W_{\bullet\bullet})$: when $W^*$ lies above the typical association strength of intermediate and longer edges, we find the behavior of Sec.~\ref{sec:timdep:results}. In turn, for lower thresholds, edges are permanently present at all spatial scales, leading to generally higher $\mathcal{C}_\bullet$ in the substrate lattice and, hence, episodic peaks of $\mathcal{C}$ since vertices in localized structures exhibit very high $\mathcal{C}_\bullet$.

\item Finally, since the considered climate networks are associated with a continuous dynamics close to the Earth's surface, high edge densities lead to very low average shortest path lengths (up to $\mathcal{L}=\mathcal{O}(1)$). This makes $\mathcal{L}$ a less informative measure at high edge densitities, even if the basic signature described in Sec.~\ref{sec:timdep:results} is still present at $\rho=0.15$.

\end{enumerate}

Conclusively, lower edge densities are generally at least as informative as higher ones, while the former rely on probably more reliable statistical associations (with respect to any kind of significance test). Information on climate dynamics becomes partially distorted at extremely low edge densities ($\rho \lesssim 0.005$), where the fraction of disconnected vertices becomes non-negligible) and blurred for extremely high edge densities ($\rho \gtrsim 0.15$).

\section{Disentangling ENSO variability} \label{sec:discussion}

Recent studies have revealed a distinct influence of ENSO variability on the topological properties of SATA networks \cite{Tsonis2008b,Yamasaki2008,Yamasaki2009,Gozolchiani2008,Gozolchiani2011,Martin2013}. Here, we study the corresponding relationship more deeply. Specifically, we hypothesize that the temporal variability of the climate network characteristics discussed in Sect.~\ref{sec:results} and \ref{sec:robustness} is mainly determined by the large-scale connectivity patterns associated with ENSO-related global climate episodes. 

The latter hypothesis is further supported by Figs.~\ref{fig:evolution1}, \ref{fig:evolution2}, \ref{fig:evolution3} and \ref{fig:extension}, where recent EN and LN episodes have been displayed for a better comparison. Here, we observe a striking coincidence between the emergence of enhanced localized structures in the SATA network and the timing of certain ENSO phases. However, a detailed inspection of these figures reveals that pronounced maxima (minima) of the different scalar network characteristics as well as the Hamming distance do not always coincide unequivocally with EN {and LN} episodes, as it was reported for another climate network approach \cite{Yamasaki2008}. For example, for the relatively strong 1990/91 EN episode, we find no marked signature in the evolving climate network characteristics. In turn, some marked extreme values of all considered measures are found in the time period{s} 1988/89 and 1992/93, which have been characterized by {a strong LN episode and the aftermath of the Mount Pinatubo eruption, respectively}. In the following, we will discuss the climatological reasons for this complex behavior and demonstrate how the signatures in different network characteristics can be utilized for disentangling the signatures of different types of {ENSO phases.

\subsection{{ENSO} vs. volcanic eruptions}

In order to understand why some time intervals display extreme values of various SATA network characteristics even without any associated ENSO phase, we first note that not only EN{/LN} episodes, but also strong volcanic eruptions have a considerable large-scale impact on the Earth's climate system \cite{Maraun2005}. To our best knowledge, a corresponding effect on climate networks has not yet been described elsewhere. Specifically, as we will explain below, both types of ``events'' can lead to the emergence of marked localized structures in the climate network. If a sufficiently large amount of aerosols is {in}jected into the stratosphere in the course of a strong volcanic eruption, it can eventually stay there for a relevant period of time (depending on the specific conditions) leading to a large-scale temporary co-evolution of SAT variability in terms of a common cooling trend over a possibly large region. Due to the corresponding relevant physical processes, this mechanism requires a certain period of time. Hence, the associated signatures in the climate network properties can only be observed with some delay. In this respect, strong volcanic eruptions can have a similar impact on the global climate system as EN{/LN} episodes in terms of a marked co{-}variation of climatic observables over a relatively large part of the globe. 

\begin{figure}
\centering
\includegraphics[width=0.485\textwidth]{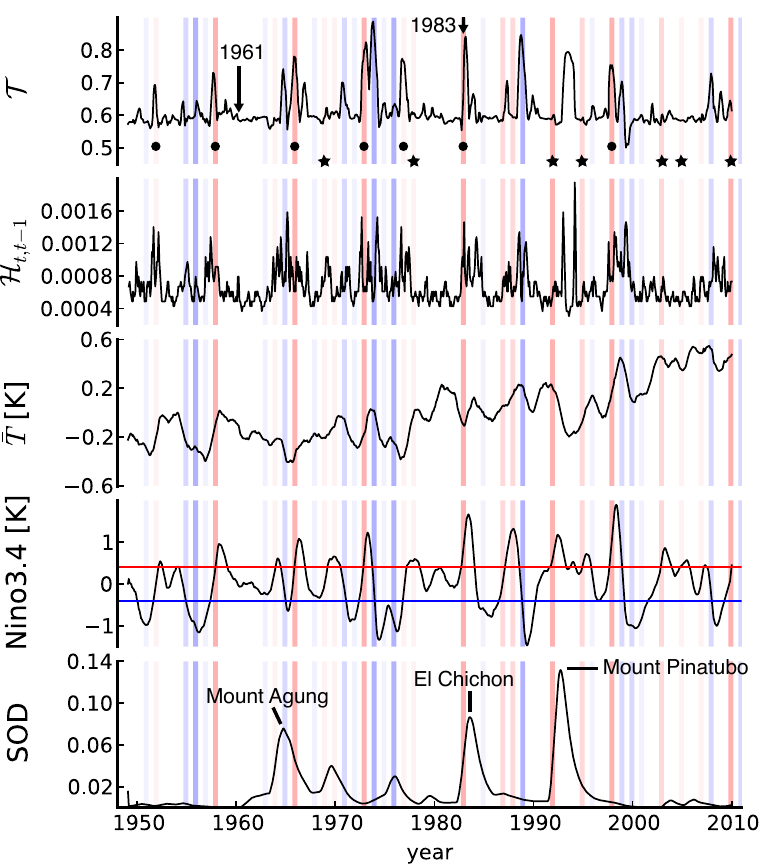}
\caption{(Color online) Evolving SATA network transitivity $\mathcal{T}$ and Hamming distance $\mathcal{H}_{t,t-1}$ for the same setting as in Fig.~\ref{fig:evolution1}. In addition, the global mean temperature anomalies $\bar{T}$, the Ni\~no 3.4 index (base period: 1948-2010, vertical lines represent thresholds of $\pm 0.4^\circ$C as an indication of EN and LN episodes~\cite{Trenberth2001}, respectively), and the stratospheric aerosol optical depth $SOD$ (i.e., the global monthly mean optical thickness at $550$ nm {wavelength}, cf.~\cite{Sato1993}) are shown. Vertical bars indicate fall and winter seasons (SON-DJF) with maximal intensity of EN (red) and LN (blue), respectively. Symbols (bullets, shifted towards the top: CT; asterisks: WP) indicate EN episodes unambigously classified in the literature (e.g., \cite{Kao2009,Kug2009}). Note that there is no 1:1 correspondence between the shown definition of EN/LN based on the Ni\~no 3.4 index and other methods.}
\label{fig:evolution4} 
\end{figure}

Figure~\ref{fig:evolution4} demonstrates that the described effect is particularly well visible in the SATA network for the year 1992/93, i.e., the time period succeeding the largest stratospheric aerosol {injection} of the 20th century, the Mount Pinatubo eruption in June 1991 \cite{McCormick1995}, which had a distinct impact on global temperatures \cite{Robock1995,Parker1996}. Like strong EN{/LN} episodes, this period has been characterized by a markedly localized structure in the climate network emerging from the Philippines and then spreading over vast parts of South-East Asia and the Western Pacific \cite{Radebach2010}, which is manifested in the climate network in terms of pronounced extreme values of all considered network quantifiers (see Figs.~\ref{fig:evolution1}, \ref{fig:evolution2}, \ref{fig:evolution3} and \ref{fig:extension}). A similar effect on the global network characteristics can be observed following the El { Chich\'on} and Mount Agung eruptions in 1982 and 1963/64, respectively, the second and third largest injections of volcanic aerosols into the stratosphere within the time interval under consideration in this work. However, the El {Chich\'on} eruption approximately coincides itself with a strong EN episode (1982/83), so that the resulting variability in the global SATA network properties cannot be unequivocally attributed to any of the two phenomena without further detailed investigations of the associated local structures. In contrast, the Mount Agung eruption clearly preceded a marked EN episode, resulting in a triple-peak signature in the evolving SATA network transitivity (Fig.~\ref{fig:evolution4}) instead of the double-peak structure exhibited by some other EN events (e.g., 1982/83 and 1997/98) or the single-peak pattern associated with the Mount Pinatubo eruption. We conjecture that this multi-peak structure highlights the emergence and disappearance of localized structures at different spatial locations relatively shortly after each other.

We emphasize that the considerable effect of strong volcanic eruptions on the global SATA network characteristics has not been studied elsewhere so far, i.e., the aforementioned results offer new directions for further research and serve as an additional proof of the usefulness of the climate network approach in general.

\subsection{El Ni\~no vs. La Ni\~na phases}

In addition to volcanic eruptions, we find that {both EN and} LN episodes can cause comparable (or even higher) peaks in the different evolving climate network properties, since they can also be associated with certain localized structures in the climate network. Specifically, Palu\v{s} \textit{et~al.} \cite{Palus2011} reported a confinement of these structure to the tropical Pacific for EN, but an extension to all tropical areas during LN phases. These findings relating to geographical aspects not explicitly studied in this work appear largely consistent with our results.

We emphasize that the physical mechanisms beyond the emergence of localized structures are, however, completely different for EN and LN episodes and volcanic eruptions. On the one hand, volcanic activity with a considerable stratospheric aerosol injection results in a consistent regional cooling trend due to reduced solar insolation, inducing generally stronger spatial correlations within a confined region. On the other hand, both extreme phases of ENSO variability (i.e., EN and LN) lead to a synchroni{z}ation of variability within large areas of the globe (due to some internal dynamics of the coupled atmosphere-ocean system \cite{Clarke2008}). As a consequence of their similar signatures in the SATA network, these different types of events apparently cannot be distinguished by just studying individual network characteristics. However, considering the temporal variability of a variety of \emph{complementary} climate network characteristics provides a more holistic picture than earlier works focusing on one parameter only \cite{Yamasaki2008,Gozolchiani2008,Yamasaki2009,Gozolchiani2011,Berezin2011,Guez2012}. Notably, this conceptual idea could be important for the general understanding of the potentials of the climate network approach. Specifically, regarding the particular problem of disentangling the signatures of ENSO variability in the evolving SATA network properties, the simultaneous study of multiple characteristics has the potential to identify some general mechanisms. Further methodological improvements such as the consideration of more sophisticated statistical association measures of time series (e.g., \cite{Pompe2011,Runge2012,Runge2012b}) remain a subject of future work.

\subsection{Discriminating different types of El Ni\~no {episodes}}

Beyond our previous considerations, recent research provided considerable evidence that there are actually two qualitatively different types of EN episodes \cite{Larkin2005,Ashok2007,Kao2009,Kug2009}. On the one hand, many EN phases follow the traditional EN pattern with strong positive sea-surface temperature anomalies starting in the {E}astern {Equatorial} Pacific and then successively {propagating westward}. This class of events particularly includes the two strongest EN episodes (1997/98 and 1982/83) recorded in the time period studied in this work with respect to the Ni\~no 3.4 index \cite{Trenberth1997, Rayner2003} as well as the 1972/73 EN episode~\cite{Kao2009,Kug2009}. On the other hand, over the last decades there has been an increasing number of EN-like phases which are characterized by large sea-surface temperature anomalies in the Central Pacific (but smaller ones in the Eastern Pacific), including the 1990/91, 1994/95, 2002/2003 and 2004/05 EN episodes. The latter type of events has been refe{r}red to as dateline El Ni{\~{n}}o~\cite{Larkin2005} or El Ni{\~{n}}o Modoki~\cite{Ashok2007} by different authors. Here, we adopt the terminology used by Kug~\textit{et~al.}~\cite{Kug2009} distinguishing between Cold Tongue (CT) and Warm Pool (WP) {episodes} corresponding to the traditional EN pattern centered in the Eastern Pacific (EP) and the Central Pacific (CP) pattern~\cite{Kao2009}, respectively.

As a novel aspect not covered in previous research, we will discuss the signatures of the two aforementioned EN types in the global SATA network characteristics next. Indeed, we are able to identify some distinctive features of the climate network associated with CT/EP {episodes}: 

First, peaks of $\mathcal{L}$, $\mathcal{T}$ and $\mathcal{H}$ perfectly coincide with the timing of the CT/EP episodes, which is \textit{not} the case for WP/CP events (cf.~Figs.~\ref{fig:evolution1} and \ref{fig:evolution4}) where such peaks are widely absent. 

Second, for strong CT/EP {episodes} the Hamming distance $\mathcal{H}$ fluctuates at a high level for more than one year, indicating a persistent redistribution (i.e., fluctuations or ``blinking'', cf.~\cite{Yamasaki2008}) of edges. For {WP/CP} episodes such fluctuations are present as well, but exhibit a considerably smaller amplitude and shorter duration. With respect to the known recent history of ENSO, we note that the three strongest unambigously classified CT/EP {episodes} (1972/73, 1982/83 and 1997/98) within the studied time interval have been directly followed by considerable LN phases (i.e., the associated Ni{\~n}o 3.4 index exhibits a particularly marked drop from strongly positive to strongly negative values, which is not as strong for WP/CP events). This sudden shift in the ENSO phase enhances the proposed mechanism of a sustained rewiring in the evolving climate networks during these periods. 

Notably, $\mathcal{L}$, $\mathcal{T}$ and $\mathcal{H}$ also show pronounced maxima {during} some strong LN episodes (this applies to both ``isolated'' LN episodes as in 1988/89 and LN phases directly following an CT/EP {episode} -- cf. the double-peak pattern of the corresponding maxima in Figs.~\ref{fig:evolution1} and \ref{fig:evolution2}), which is due to the emergence of (though a possibly different type of) localized structures in the climate network. {However, for other LN phases, such peaks are absent. This observation suggests the existence of two different climatological mechanisms, which could result in a classification of LN episodes in a similar way as for EN phases. We leave a more detailed investigation and discussion of this idea for future work.}

The distinctively different behavior of SATA networks during the two types of EN episodes can probably be explained by reconsidering the {associated} typical spatio-temporal patterns. On the one hand, CT/EP episodes exhibit a {relatively} sharp, regionally confined pattern leading to a common SAT trend starting from the Eastern Equatorial Pacific and then {propagating} westward. In this spirit, the spatio-temporal signature in the SAT field resembles a wave travelling through the Equatorial Pacific from East to West, leading to a successive synchronization of tropical climate variability over an increasingly large region. On the other hand, the typical pattern of WP/CP {episodes} commonly appears like a diffuse pulse spreading from one region in the Central Equatorial Pacific into different directions. The associated EOF patterns display more fuzzy spatial structures and are less well localized in space than those of CT/EP episodes \cite{Ashok2007,Kug2009,Kao2009}. Due to this spatio-temporal footprint, the spatial correlations in the Equatorial Pacific change their magnitude in a much less coordinated and marked way than under the influence of the sharper pattern associated with CT/EP episodes. As a consequence, there is only a relatively minor redistribution of connectivity in the SATA network, explaining the weaker signatures in the considered network characteristics. 

Notably, {at this point} we are not able to give a complete classification of EN episodes based on our complex network characteristics. This is particularly due to the fact that there are EN episodes of mixed characteristics already known in the literature, so that there is no unique reference for classification. Moreover, the definition of EN itself is partly ambiguous and depends on the specific method or index of choice (e.g.,  \cite{Kiladis1988,Trenberth1997,Larkin2005,Ashok2007,Kug2009,Hendon2009,Yeh2009,Kim2009,McPhaden2009,Kim2011,Hu2012,Graf2012,Miyakoda2012}).

\subsection{Network characteristics and climate-related indices}

In order to further support our previous results, the cross-correlation functions between the evolving climate network characteristics $\mathcal{T}$ and $\mathcal{H}_{t,t-1}$ on the one hand, and the global average temperature anomalies $\bar{T}$, the absolute value of the Ni\~no 3.4 index as well as the stratospheric aerosol optical depth $SOD$ (as indicators of ENSO and volcanic activity, respectively) on the other hand have been computed (Fig.~\ref{fig:correlations2}). Notably, we do not find any systematic effect of the average temperature anomalies on the evolving climate networks. This indicates that the general global warming trend is not directly reflected in the corresponding network properties. However, such trends are practically only relevant for time scales clearly above the window sizes of one year studied in this work. In turn, dynamical characteristics such as captured by evolving climate network analysis reveal signatures that go clearly beyond the behavior of global mean temperatures.

\begin{figure}
\centering
\includegraphics[width=0.485\textwidth]{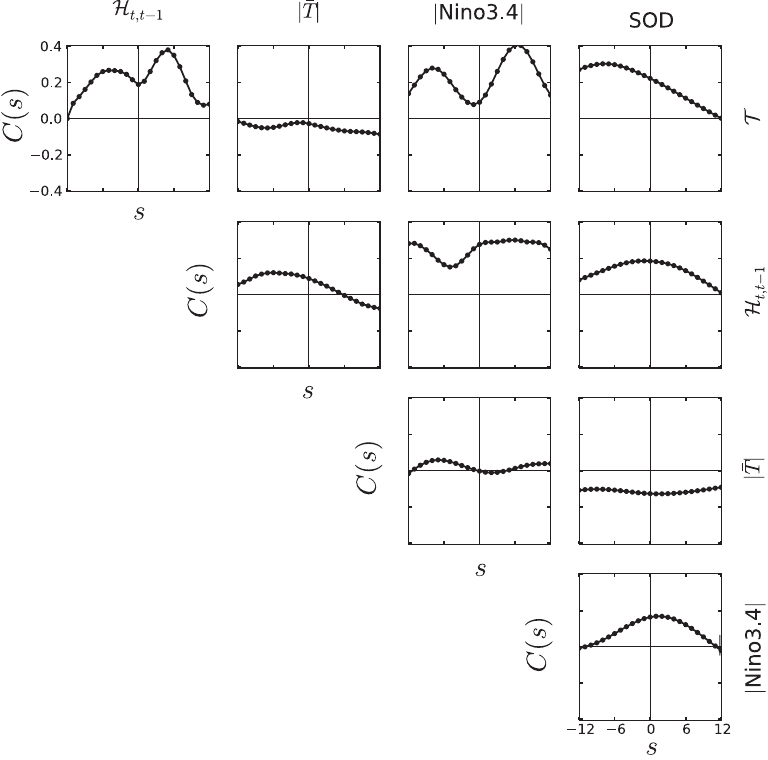}
\caption{Linear cross-correlation functions $C({s})$ between the evolving SATA network characteristics transitivity $\mathcal{T}$ and Hamming distance $\mathcal{H}_{t,t-1}$ (obtained based on lag-zero cross-correlation with fixed $\rho=0.005$), {the} global mean temperature anomaly $\bar{T}$, the absolute value of {the} Ni\~no 3.4 index, and {the} $SOD$ index.}
\label{fig:correlations2} 
\end{figure}

Regarding the impact of volcanic activity, the obtained results demonstrate a considerable influence on the SATA network topology. Despite recent findings suggesting a possible {effect} of climatic processes on volcanic activity on longer time scales~\cite{Kutterolf2013,McGuire2010}, we can practically rule out a significant climatic forcing of volcanism at the time scales considered in this work. Specifically, the network transitivity shows similar variations as the $SOD$ index with a delay of about 8 months, which is in reasonable agreement with the typical lifetime of volcanic particles in the stratosphere and the expected delay of SAT changes due to a reduction of solar irradiation. Note, however, that the $SOD$ index only shows some distinct events and remains close to 0 for most of the time. In this case, linear cross-correlation is not the best-suited measure for characterizing the co-occurrence of volcanic eruptions and peaks in the SATA network characterics. In contrast, event-based characteristics such as event synchronization \cite{QuianQuiroga2002,Malik2010,Malik2011} or coincidence analysis \cite{Donges2011PNAS} are tailored for such purposes, but require a larger number of events than those recorded in the studied data sets.

More interestingly, cross-correlation analysis reveals considerable interdependences between the SATA network properties and the absolute value of the Ni{\~n}o 3.4 index, characterizing the deviation of sea-surface temperature anomalies from the standard values in some defined region of the Pacific associated with ENSO. From Fig.~\ref{fig:correlations2}, we find that network transitivity $\mathcal{T}$ and Hamming distance $\mathcal{H}_{t,t-1}$ show distinct maxima preceding the peak amplitude of EN and LN by 6 and 5 months, respectively, with correlation values of about 0.4. Moreover, secondary maxima of the cross-correlation functions with smaller amplitude are found 8 and 5 months after the correspondent peaks {of the ENSO index}, respectively. While we can unequivocally attribute this finding for the Hamming distance as resulting from the largest rate of redistribution of connectivity within the SATA network, the corresponding signature of $\mathcal{T}$ seems to rather relate to the presence of common SAT trends over a substantial region associated with both the emergence and disappearance of localized structures in the Equatorial Pacific and beyond. (In turn, for the $SOD$ index there exists only one maximum, indicating that the disappearance of the associated characteristic structures in the SAT field behaves fundamentally different than for EN{/LN} episodes.) It will be subject of future studies to what extent this information, in combination with the distinct temporal variability profiles of different SATA network measures, does not only provide important novel insights into the function of the climate system in general, but can be specifically exploited for anticipating or even predicting type and strength of approaching EN and LN episodes~\cite{Ludescher2013}.

\section{Conclusions} \label{sec:conclusions}

In this work, we have introduced a novel viewpoint on inter-annual climate variability in terms of evolving climate network analysis, i.e., studying the variation of a set of complementary global climate network characteristics with time. Our analysis has provided deep new insights into the functionality of the global climate system and impacts of different types of climate episodes, particularly such related to ENSO variability. Our findings particularly highlight the effect of ``classical'' El Ni{\~{n}}o {(EN) and some La Ni\~{n}a (LN)} episodes as well as very strong volcanic eruptions on the global climate system. Specifically, {all three} types of ``events'' lead to a common large-scale temperature trend (i.e., some kind of synchronization) over a considerably large region. In the climate network, this results in the emergence of marked localized structures, for ENSO particularly in the tropical Pacific. Note that the results presented in this work go significantly beyond those of previous research. As a main new achievement, we can {not only} clearly distinguish between the signatures of different {ENSO phases, but also differentiate different types of EN and LN episodes,} which has not yet been possible by other climate network approaches. {Specifically}, we have developed some initial understanding of similarities and differences between the climate network reflections of physical mechanisms acting during strong volcanic eruptions and different types of EN {and LN} episodes. 

Beyond the specific consideration of ENSO variability, our results have led to a substantially improved understanding of the structures present in climate networks based on surface air temperatures. As a general finding, we have proposed a new simple conceptual view of the climate network based on the alternating presence of different types of structures: the substrate lattice mainly capturing short-range connections versus enhanced localized structures (i.e., densely connected parts of the network covering larger spatial scales). In this respect, the temporal variability of {the climate} network topology can be understood as an effect of a persistent redistribution of connectivity between these different types of structures.

We note that our approach is distinctively different from those previously used by other authors in the sense that we have considered a multiplicity of comprehensive measures from complex network theory. Only this consideration of complementary characteristics allowed deriving a holistic understanding of the underlying dynamical processes. Motivated by its successful application, we suggest further using not only the global characteristics of climate networks as studied in this work, but also the associated spatial patterns of (both topological and geometric) vertex properties (i.e., information on the placement of edges in physical space) for future investigations on the detailed spatial backbones of different climate episodes. Initial results in this direction can be found in \cite{Radebach2010}. We conjecture that this evolving network approach has great potentials for supplementing other studies based on traditional methods of multivariate statistics such as EOF analysis.  

As underlined by our analysis, different ENSO phases have a distinct impact on the spatial organization of the global climate system. Since ENSO is a coupled atmosphere-ocean phenomenon, we additionally suggest the consideration of complementary climatological observables (e.g., geopotential height, sea-surface temperatures, sea-level pressure, etc.) in corresponding future analyses. A methodological extension that is particularly tailored for such investigations are coupled climate networks~\cite{Donges2011,Feng2012}.

\begin{acknowledgements}
This work has been funded by the Leibniz association (project ECONS), the Federal Ministry for Education and Research (BMBF) via the Potsdam Research Cluster for Georisk Analysis, Environmental Change and Sustainability (PROGRESS), the German Research Foundation (DFG project ``Interactions and complex structures in the dynamics of changing climate''), the German Federal Environmental Agency, and the EU FP7 projects SUMO, LINC and MEDIATION. AR, JR and JFD acknowledge financial support of the German National Academic Foundation.  JR was partially supported by the German Environmental Foundation. NCEP {r}eanalysis data have been provided by the NOAA/OAR/ESRL PSD, Boulder, Colorado, USA, from their {w}ebsite at \texttt{http://www.esrl.noaa.gov/psd/}. We thank Norbert Marwan and Kira Rehfeld for inspiring discussions and support, and Roger Grzondziel and Ciaron Linstead for help with the IBM iDataPlex Cluster at the Potsdam Institute for Climate Impact Research. Complex network measures have been calculated using the software packages \texttt{igraph} \cite{Csardi2006} and \texttt{pyunicorn} {\cite{donges2013advanced}}. Finally, we are indebted to two anonymous referees for their insightful comments on an earlier version of this paper.
\end{acknowledgements}

\bibliographystyle{apsrev}      
\bibliography{npg_alex_rev}

\end{document}